\documentclass{aa}

\usepackage{graphicx,amsmath,txfonts}

\begin{document}
\newcommand{\mearth}{M_\oplus}
\title{
Two phase, inward-then-outward migration of Jupiter and Saturn in the gaseous Solar Nebula
}
\author{Arnaud Pierens \inst{1,2} \and Sean~N. Raymond \inst{1,2} }
\authorrunning{Pierens \& Raymond}
\titlerunning{Two-phase migration of Jupiter and Saturn}
\institute{Universit\'e de Bordeaux, Observatoire Aquitain des Sciences de l'Univers,
    BP89 33271 Floirac Cedex, France \label{inst1} \and
   CNRS, UMR 5804, Laboratoire d'Astrophysique de Bordeaux,
    BP89 33271 Floirac Cedex, France \label{inst2} \\
     \email{arnaud.pierens@obs.u-bordeaux1.fr}}

\date{}

\abstract
{
It has recently been shown that the terrestrial planets and asteroid belt can be reproduced if the giant planets underwent an inward-then-outward migration (the "Grand Tack"; Walsh et al 2011).  Inward migration occurs when Jupiter opens a gap and type II migrates inward.  The planets "tack" and migrate outward when Saturn reaches the gap-opening mass and is caught in the 3:2 resonance with Jupiter.}
{The aim is to test the viability of the Grand Tack model and  to study the dynamical evolution of Jupiter and Saturn during their growth from $10$ $\mearth$ cores.}
{We have performed numerical simulations using a grid--based hydrodynamical code. Most of our simulations assume an isothermal equation of state for the disk but a subset use a fully-radiative version of the code.}
{For an isothermal disk the two phase migration of Jupiter and Saturn is very robust and independent of the mass-growth history of these planets provided the disk is cool enough.  For a radiative disk the we find some outcomes with two phase migrations and others with more complicated behavior.  We construct a simple, 1-D model of an evolving viscous disk to calculate the evolution of the disk's radiative properties: the disk transitions from radiative to isothermal from its outermost regions inward in time.}
{We show that a two-phase migration is a natural outcome at late times even under the limiting assumption that isothermal conditions are required.  Thus, our simulations provide strong support for the Grand Tack scenario. 
}

\maketitle

\section{Introduction}

The current paradigm of the origin and evolution of the Solar System's giant planets follows several distinct stages:
\begin{enumerate}
\item The cores of Jupiter and Saturn form by accretion of planetesimals (e.g., Kokubo \& Ida 1998; Levison et al. 2010).  The timescales of accretion are poorly constrained because as they grow the cores migrate due to both the back-reaction from planetesimal scattering (Fernandez \& Ip 1984; Kirsh et al. 2009) and type I (tidal) interactions with the gaseous protoplanetary disk (Goldreich \& Tremaine 1980; Ward 1986; Paardekooper et al. 2010).  
\item Jupiter and Saturn's cores slowly accrete gas and each undergo a phase of rapid gas accretion (e.g., Mizuno 1980, Pollack et al. 1996).  The rapid phase of accretion is triggered when the mass in each planet's gaseous envelope is comparable to the core mass.  Runaway gas accretion lasts for roughly the local Kelvin-Helmholtz time and ceases when the planet opens an annular gap in the disk and transitions to type II migration (Lin \& Papaloizou 1986; Ward 1997).  Because of its larger mass and smaller orbital radius, Jupiter is thought to have undergone runaway gas accretion before Saturn.
\item Once fully-formed, Saturn migrated faster (Masset \& Papaloizou 2003), caught up to Jupiter, and was trapped in 3:2 resonance (Pierens \& Nelson 2008).  Interestingly, this result is found to be a very robust outcome of the simulations, independent on the earlier evolution of Saturn's core. For instance, Pierens \& Nelson (2008) investigated the scenario in which Saturn's core is initially trapped at the edge of Jupiter's gap and grows through  gas accretion from the disk. In that case, they demonstrated that although Saturn is temporarily locked in the 2:1 resonance with Jupiter, it becomes ultimately trapped in the 3:2 resonance. 
\item Once Jupiter and Saturn are trapped in 3:2 resonance, the gaps carved by the two planets in the Solar Nebula overlap.  Saturn's gap is not as deep as Jupiter's (due to its smaller mass), and this causes Jupiter and Saturn to migrate outward while remaining in 3:2 resonance, provided that both the disk thickness and the disk viscosity are small enough (Masset \& Snellgrove 2001; Morbidelli \& Crida 2007).   Outward migration is stopped when the disk dissipates  or, if the disk is flared, at a critical distance where the disk is too thick and the structure of the two planets' common gap is compromised (Crida et al 2009).  \footnote{We note that this step is somewhat uncertain because both the migration and accretion rates of giant planet cores should a roughly linear dependence on the planet mass.  Thus, we naively expect that Saturn's gas accretion should mimic Jupiter's as it migrates inward, meaning that Saturn should be roughly $1 M_J$ when it catches up to Jupiter, precluding outward migration, which requires a Saturn/Jupiter mass ratio of roughly 1/2 or smaller (Masset \& Snellgrove 2001).  The solution to this problem is not clear: it may involve a change in the disk opacity to allow Saturn's rapid type III migration to last for longer than Jupiter's.  Such rapid migration has been invoked to explain the 3:2 resonant exoplanet system HD 45364 (Rein et al. 2010).  Of course it is reasonable to expect that this mechanism is probably not universal, since it depends on details such as the timing of core formation and the disk properties.  Thus, in many exoplanet systems "Saturn" would have reached $1 M_J$ and the two planets would not tacked and migrated outward.  The architecture of such systems naturally would not resemble the Solar System.  Understanding the statistical distribution of giant exoplanetary systems can therefore place constraints on their early evolution and the frequency of "grand tacks". } 
\item After the dissipation of the gas disk, planetesimal-driven migration causes a large-scale spreading of the planets' orbits because Jupiter is the only planet for which the ejection of small bodies is more probable than inward scattering (Fernandez \& Ip 1984; Hahn \& Malhotra 1999).  In the "Nice model", Jupiter and Saturn are assumed to have formed interior to their mutual 2:1 resonance and, when they cross it, an instability is triggered that causes the Late Heavy Bombardment (Gomes et al. 2005).  Recent work has shown that the Nice model is still valid if more realistic initial conditions are used, with Jupiter and Saturn in 3:2 resonance and Uranus and Neptune also trapped in a resonant chain (Morbidelli et al. 2007; Batygin \& Brown 2010).   The Nice model can reproduce the giant planets' final orbits (Tsiganis et al. 2005), the orbital distribution of Jupiter's Trojan asteroids (Morbidelli et al. 2005), and several other characteristics of the Solar System's small body populations.  
\end{enumerate}

Although the detailed orbital evolution of the giant planets is not known, these steps explain their origin in broad strokes.  By assembling steps 2-4, Walsh et al. (2011) recently proposed a new model to explain the origin of the inner Solar System called the "Grand Tack".  In this model, Jupiter formed at $\sim 2-5$ AU, migrated inward then "tacked" (i.e., changed the direction of its migration) at an orbital distance of $\sim$1.5 AU when Saturn caught up and was trapped in 3:2 resonance and migrated back out past 5 AU.  Jupiter's tack at 1.5 AU truncates the inner disk of planetary embryos and planetesimals from which the terrestrial planets formed at about 1 AU.  This type of narrow truncated disk represents the only initial conditions known to satisfactorily reproduce the terrestrial planets, in particular the small mass of Mars compared with Earth (Wetherill 1978; Hansen 2009; Raymond et al. 2009; Walsh et al. 2011).  An additional success of the Grand Tack model is that the asteroid belt is naturally repopulated from two distinct populations corresponding to the C- and S- type asteroids.  At the end of the Grand Tack, the giant planets' orbits represent the initial conditions for the Nice model (Raymond et al., in prep.).  

The goal of this paper is to test the viability of the evolution of Jupiter and Saturn in the Grand Tack model (Walsh et al 2011).  To accomplish this we use the GENESIS hydrocode to simulate the growth and migration of Jupiter and Saturn from $10$ $\mearth$ cores.  With respect to previous simulations (Masset \& Snellgrove 2001; Morbidelli \& Crida 2007; Pierens \& Nelson 2008), we consider a self-consistent scenario in which the cores of Jupiter and Saturn slowly grow to full-fledged gas giants by accreting gas from the disk.  We find that a two phase migration of Jupiter and Saturn is a very robust outcome in isothermal disks, but occurs in only one of two simulations in radiative disks.  We place our simulations in the context of the evolving Solar Nebula using a 1-D diffusion algorithm that differentiates between radiative and isothermal behavior.  Our results strongly favor a two-phase migration of Jupiter and Saturn, and support the Grand Tack. 

The paper is organized as follows. In Sect. 2, we describe the hydrodynamical model. In Sect. 3,
we present the results of isothermal simulations.  In Sect. 4 we present results of radiative simulations.  In Sect. 5 we construct a 1-D model of the Solar Nebula to show when the disk should be isothermal or radiative.  Finally, we discuss our results and draw conclusions in Sect. 6.

%

\section{The hydrodynamical model}

\subsection{Numerical method}

In this paper, we adopt a 2D disk model for which all the physical 
quantities are vertically averaged. We work in a non-rotating frame,
and adopt cylindrical polar coordinates $(R,\phi)$ with the origin
located at the position of the central star. Indirect terms resulting
from the fact that this frame is non-inertial are incorporated in the
equations governing the disk evolution (e.g. Nelson et al. 2000). These 
are solved using the GENESIS hydrocode for which a full description can
be found for example in De Val-Borro et al. (2006). The evolution of each
 planetary orbit is computed using a fifth-order Runge-Kutta integrator
(Press et al. 1992) and by calculating the torques exerted by the disk
on each planet. The disk material located inside the Hill 
sphere of each planet is excluded when computing 
the disk torques (but see \S below). We also employ a softening parameter $b=0.6 H$
,where $H$ is the disk scale height, when calculating the planet potentials.

In the simulations presented here, we use $N_R=608$ radial grid cells
uniformly distributed between $R_{in}=0.25$ and $R_{out}=7$ and 
$N_{\phi}=700$ azimuthal grid cells uniformly distributed between 
$\phi_{min}=0$ and $\phi_{max}=2\pi$. Wave-killing zones are employed
 for $R<0.5$ and $R>6.5$ in order to avoid wave reflections at the disk edges 
(de Val-Borro et al. 2006).

For most of the simulations, we adopt a locally isothermal equation of state with a fixed
temperature profile given by $T=T_0(R/R_0)^{-\beta}$ where $\beta=1$ and where 
$T_0$ is the temperature at $R_0=1$.  This corresponds to a disk with constant 
aspect ratio $h$ for which we consider values of $h=0.03$, $0.04$, $0.05$. The 
initial surface density profile is chosen to be $\Sigma(R)=\Sigma_0(R/R_0)^{-\sigma}$ 
with $\sigma=1/2$, $3/2$ and $\Sigma_0=4\times 10^{-4}$.  In our units, this 
corresponds to a disk containing $\sim 0.04$ $M_\odot$ within $40$ AU. 

The adopted computational units are such that the mass of the central star 
$M_*=1$ corresponds to one Solar mass, the gravitational constant is 
$G=1$ and the radius $R=1$ in the computational domain corresponds to $5$ AU.  Thus, in the rest of the paper time is measured in units of orbital periods at $R=1$ ($\sim 10$ years for 5 AU).  However, we note that these simulations can be scaled to different disk parameters.  For example, if we assume $R=1$ corresponds to 1 AU, then the corresponding disk mass is roughly 10 times larger (for our fiducial $R^{-1/2}$ surface density profile).  To run simulations that better reproduce the conditions at $\sim$ 1 AU would require integration times that are ten times longer with disks that are one tenth the density, and are not computationally feasible given our current resources.  Thus, our simulations are intended to demonstrate the relevant mechanisms in a similar setting that is, admittedly, somewhat more distant.  However, we did perform one simulation in a very low-mass disk in which $R=1$ corresponded to 1 AU.  That simulation, presented briefly in Sect. 5, serves to validate our results.

 We have also performed a few additional radiative runs where the thermal energy is
solved. Source terms corresponding to viscous heating and local radiative cooling from the disk 
surfaces are implemented and handled 
similarly to Kley \& Crida (2008), except that we use the Rosseland mean opacity given by Bell \& Lin (1994). 
Heat diffusion in the disk midplane is not taken into account in the simulations presented here.

Viscous stresses probably arising from MHD turbulence are modelled using the 
standard 'alpha' prescription for the disk viscosity $\nu=\alpha c_s H$ 
(Shakura \& Sunyaev 1973), where $c_s$ is the isothermal sound speed and $H$ 
is the disk scale height. In the simulations presented here, we use values of 
$\alpha=2\times 10^{-3}$ and $\alpha=4\times 10^{-3}$.

\begin{table*}
\centering{}
\begin{tabular} {c c c c c c c c c}
\hline \hline
Model & $m_{J,i}(\mearth)$ & $m_{S,i}(\mearth)$  & $f_J$ & $f_S$ & $x_J$ & $h$ &  $\alpha$ & $\sigma$ \\
\hline
$I1$ & $10$ & $10$ & $5/3$ & $5/3$ & $0.5$ & $0.04$  & $2\times 10^{-3}$ & $1/2$ \\
$I2$ & $10$ & $10$ & $5/3$ & $5/6$ & $0.5$ & $0.04$  & $2\times 10^{-3}$ & $1/2$\\
$I3$ & $10$ & $10$ & $5/3$ & $5/3$ & $0$ & $0.04$  & $2\times 10^{-3}$ & $1/2$ \\
$I4$ & $10$ & $10$ & $5/3$ & $5/3$ & $1$ & $0.04$  & $2\times 10^{-3}$ & $1/2$\\
$I5$ & $10$ & $10$ & $5/3$ & $5/3$ & $0.5$ & $0.04$  & $4\times 10^{-3}$ & $1/2$ \\
$I6$ & $10$ & $10$ & $5/3$ & $5/3$ & $0.5$ & $0.03$  & $2\times 10^{-3}$ & $1/2$ \\
$I7$ & $10$ & $10$ & $5/3$ & $5/3$ & $0.5$ & $0.05$  & $2\times 10^{-3}$ & $1/2$ \\
$I8$ & $10$ & $10$ & $5/3$ & $5/3$ & $0.5$ & $0.04$  & $2\times 10^{-3}$ & $3/2$ \\
$I9$ & $10$ & $10$ & $5/3$ & $5/3$ & $1$ & $0.04$  & $2\times 10^{-3}$ & $3/2$ \\
$R1$ & $10$ & $10$ & $5/3$ & $5/3$ & $0.5$ & $rad$  & $2\times 10^{-3}$ & $1/2$ \\
$R2$ & $10$ & $10$ & $5/3$ & $5/3$ & $0$ & $rad$  & $2\times 10^{-3}$ & $1/2$ \\
\hline
\end{tabular}
\caption{Parameters used in the simulations.}
\label{table1}
\end{table*}

\subsection{Initial conditions}

 In our simulations, the masses of Jupiter $m_J$ and Saturn $m_S$ are initiated with values of 
$m_{J,i}=m_{S,i}=10$ $\mearth$; and the cores of Jupiter and Saturn are placed on circular orbits at $a_J=2$ and $a_S=2.65$, just exterior to their mutual 3:2 mean motion resonance.

In an isothermal disk, the type I migration timescale of a planet with mass $m_p$, semimajor axis $a_p$  and on a circular orbit with angular frequency $\Omega_p$ can be estimated by 
(Paardekooper et al. 2010):
\begin{equation}
\tau_{mig}=(1.6+\beta+0.7\sigma)^{-1} \frac{M_\star}{m_p}\frac{M_\star}
{\Sigma(a_p) a_p^2}h^2\Omega_p^{-1}.
\label{eq:taumig}
\end{equation}
Because of Eq. \ref{eq:taumig}, we expect Jupiter and Saturn's cores, embedded in a disk model with $\sigma<3/2$ , to  
undergo convergent migration and become eventually trapped in the $3:2$ resonance.  In contrast, a disk model with $\sigma > 3/2$ should lead to divergent migration. 

Once the cores have evolved for $\sim 500$ orbits of the innermost embryo, we allow 
Jupiter's core to accrete gas from the disk.  For each timestep $\Delta t$, accretion is modeled by reducing the surface density in the grid cells located within a 
distance $R_{acc}$ of the planet by a factor $1 -f_J \Delta t$. Following Paardekooper 
\& Mellema (2008), we set $f_J=5/3$ in our simulations. Furthermore, we choose 
$R_{acc}=0.1$ $R_{H,J}$ where $R_{H,J}$ is the Hill radius of the planet ($R_{H,J} = a_J 
\left(m_J/3 M_\odot \right)^{1/3}$); this value is small enough to ensure that the accretion procedure is independent of our choice of $f_J$ (Tanigawa \& Watanabe 2002). 

Accretion onto Saturn's core is handled in the same way but we use different values for the accretion parameter $f_S$. In this work, this process is switched on when Jupiter has grown to a fraction $x_J=m_J/M_J$ of its final mass $M_J=318$ $\mearth$. We varied the value for $x_J$ in such a way that different starting times for accretion 
onto Saturn's core are considered. For instance, we used $x_J=0.5$ in most of the simulations presented here, meaning that accretion onto Saturn's core is triggered once Jupiter has reached half of its final mass.  We also tested values for $x_J$ of 0 (concurrent accretion of both cores) and 1 (isolated growth of each core in succession).  

The parameters for all of our locally isothermal simulations are shown in Table \ref{table1}.  Our fiducial simulation (model I1) had $h=0.04$, $\sigma=0.5$, $\alpha=2\times 10^{-3}$, $f_J = f_S = 5/3$, and $\sigma = 1/2$.   We performed additional runs varying the disk aspect ratio, viscosity, and surface density profile, varying each parameter orthogonally to our fiducial case and keeping the other parameters fixed. 

\section{Isothermal simulations}
\subsection{Evolution of our fiducial case}
\label{base}

\begin{figure*}
\centering
\includegraphics[width=0.9\textwidth]{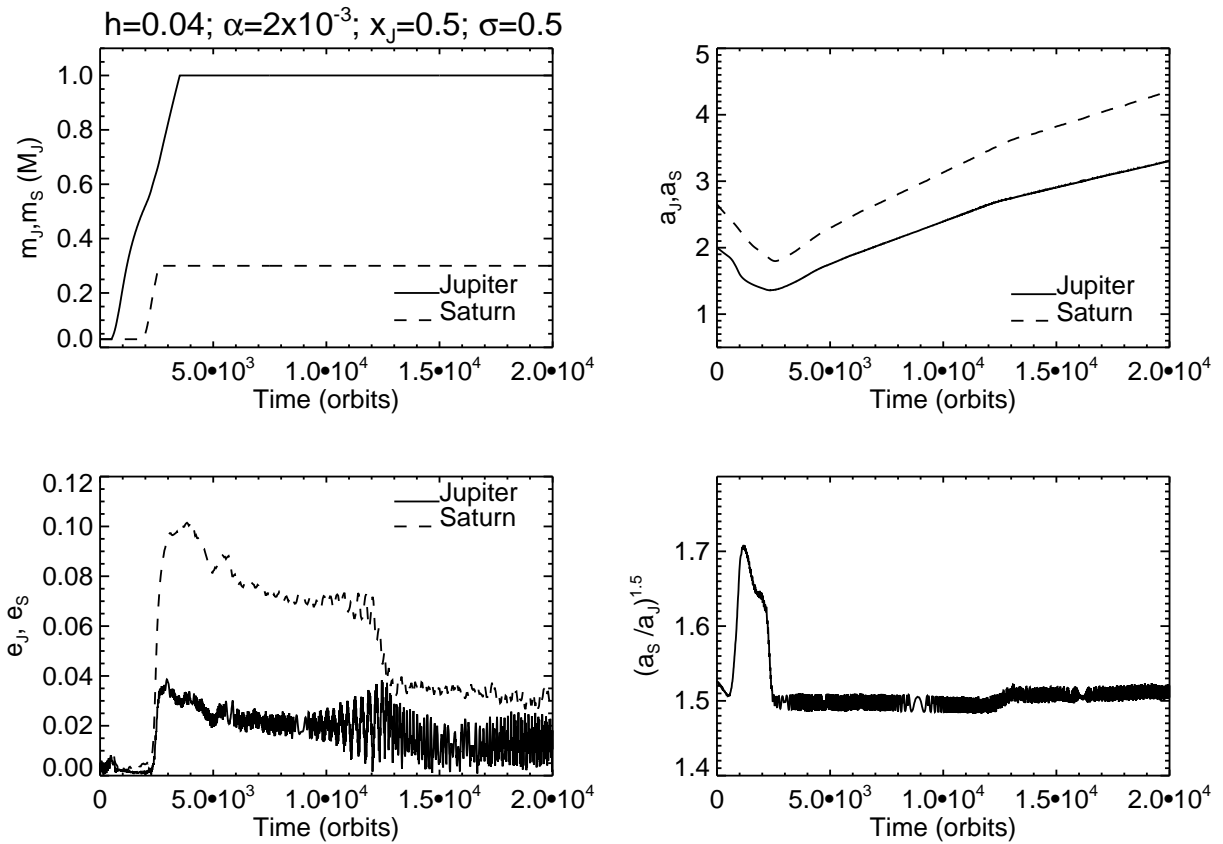}
\caption{{\it Upper left (first) panel:}  the time evolution of the masses of 
 Jupiter (solid line) and Saturn (dashed line) for run I1. 
{\it Upper right (second) panel:} time evolution of the semimajor axes. {\it Third panel:} 
time evolution of the eccentricities. {\it Fourth panel:} time evolution of the period ratio 
$(a_S/a_J)^{1.5}$.}
\label{orbits}
\end{figure*}
  
Fig. \ref{orbits} shows the evolution of our fiducial model I1 in which Saturn starts to
accrete gas once Jupiter has grown to half of its final mass. The time evolution of the planet masses is shown in the first panel. It is worth noting that the gas accretion rate onto Saturn's core is actually higher than Jupiter's. This is simply because the disk mass included in the feeding zone of a planet with semimajor axis $a_p$ increases as $\sim a_p^{2-\sigma}$. 
The second panel depicts the evolution of the semimajor axes. 
In this run, the two cores migrate convergently  at early times and are captured 
in the 3:2 resonance.  The resonance is maintained until
$t\sim 500$ orbits when Jupiter starts to accrete gas from the disk. Then, the two cores move away from each other because Jupiter's migration rate increases as $\sim m_J$ before it opens a gap. This has the consequence 
of disrupting the 3:2 resonance.  At later times, the migration once again becomes convergent as Jupiter clears a gap and transitions to slower, type II migration, while Saturn starts accreting gas and its type I migration accelerates. 
The fourth panel of Fig. \ref{orbits} shows the evolution of the period ratio $(a_S/a_J)^{3/2}$. The slowing down of Jupiter's migration due to the onset of non-linear effects (once $m_J\lesssim 0.3 M_J$) is illustrated by the presence of a local maximum in the period ratio at $t\sim 1200$ orbits,
while the onset of gas accretion onto Saturn's core is responsible for the sudden drop in period ratio at $t\sim 2000$ orbits. 

\begin{figure}
\centering
\includegraphics[width=0.98\columnwidth]{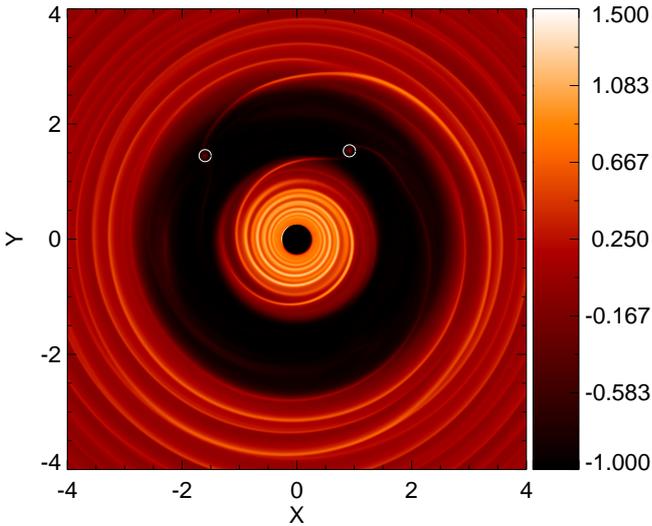}
\caption{Snapshot of the perturbed disk surface density for model I1 at a point in time where Jupiter and Saturn are 
fully formed, locked in a 3:2 MMR and migrate outward. }
\label{fig:disk2d_i1}
\end{figure}

Convergent migration causes the 3:2 resonance to be recovered at $t\sim 2300$ orbits and 
the resonance remains stable for the duration of the simulation. This is illustrated by the upper panel 
of Fig. \ref{fig:angles} which shows the time evolution of the two resonant angles associated with the 3:2 resonance:
\begin{equation}
\psi=3\lambda_S-2\lambda_J-\omega_J \quad {\rm and}\quad \phi=3\lambda_S-2\lambda_J-\omega_S,
\end{equation}
where $\lambda_J$ ($\lambda_S$) 
and $\omega_J$ ($\omega_S$) are the mean longitude and longitude of pericentre of Jupiter (and Saturn).  Capture in the 3:2 resonance causes the eccentricities of both planets to grow 
to $e_J\sim 0.03$ and $e_S\sim 0.1$, as seen in the third
panel in Fig. \ref{orbits}. In agreement with previous studies
(Masset \& Snellgrove 2001, Morbidelli \& Crida 2007, Pierens \& Nelson 2008), the long-term outcome for model I1 after capture in the 3:2 resonance is outward migration of the Jupiter-Saturn system with the two planets 
maintaining the 3:2 resonance and sharing a common gap.  Here, the migration reversal occurs at $t\sim 2500$ orbits, when $m_J\sim 0.6$ $M_J$ and $m_S \sim 0.3$ $M_J$.  
These values are in reasonable agreement with Masset \& Snellgrove (2001) who estimated a critical mass ratio of 
$m_s/m_J \lesssim 0.62$ for the positive torque exerted by the inner disk on Jupiter to be larger than the negative 
torque exerted by the outer disk on Saturn. 
 Fig.~\ref{fig:disk2d_i1} shows a snapshot of the disk at a point in time  where  Jupiter and Saturn are fully-formed, locked in 3:2 MMR and migrate outward.  

For this model, we observe a trend for the amplitude of the resonant angles to slightly increase with time to such an extent that $\phi$ switches from libration to circulation at $t\sim 1.3\times 10^4$ orbits 
(see upper panel of Fig. \ref{fig:angles}). This causes not only the
outward migration rate to subsequently slow down (second panel) but also the eccentricities to
slightly decrease to values such that $e_J\sim 0.02$ and $e_S\sim 0.04$ at the end of the simulation (third panel).  It is interesting to note that some of the known exoplanet systems appear to exhibit libration of only one resonant angle. In the case of HD128311 (Vogt et al 2005), this behavior might be explained by a scattering event (Sandor \& Kley 2006) or by effects due to turbulence (Rein \& Papaloizou 2009). 

In simulation I1, the change from libration to circulation of one resonant angle occurs when Saturn is at $\sim 3.5$ numerical units, roughly half the distance of the outer disk edge.  To test whether this could have been a numerical effect caused by unresolved high-order density waves (perhaps associated with the outer 3:1 resonance with Saturn) we re-ran the latter evolution of simulation I1 but with an outer disk boundary at 5 numerical units rather than 7.  In the test simulation, the change from libration to circulation again occurred when Saturn was at $\sim 3.5$ units, showing that the outer disk was not the cause of the change.  

It appears that  one possibility that caused the shift from libration to circulation of the resonant angle was the corotation torque exerted on Saturn. Examination of the torques exerted on Saturn indeed reveals a tendency for the torque exerted by the disk material located in between the two planets to increase. This suggests that, as the orbital separation between Jupiter and Saturn increases during their outward migration in 3:2 resonance, gas flowing across Saturn's orbit can feed the common gap and consequently exert a positive horseshoe drag on Saturn. Such a process tends to further push Saturn outward (Zhang \& Zhou 2010), which may subsequently lead to the disruption of the apsidal corotation.  This effect may have important consequences for the ability of two planets to undergo long-range outward migration via the Masset \& Snellgrove (2001) mechanism.

\subsection{Effect of simulation parameters}

 In the following, we investigate how the evolution of the system depends on the simulations parameters. We test the effect 
of varying the start time of Saturn's growth, disk aspect ratio, viscosity and surface density profile. 
Also not shown here, we have also tested the effect of changing the accretion parameter $f_S$ of Saturn and have performed
one simulation with $f_S=5/6$ (model I2). In that case, the evolution of the system was found to be very similar to that 
obtained in model I1. 

\subsubsection{Dependence on the start time of Saturn's growth}
\label{sec:varyxj}
In order to examine how the evolution depends on the mass-growth history of Jupiter and Saturn, we performed two additional simulations varying the time when Saturn's core starts to accrete gas from the disc (Table 1). 
In simulation I3, accretion onto the cores of Jupiter and Saturn are switched on at the same time while in simulation I4 Saturn's accretion started once Jupiter was fully formed.
Fig. \ref{fig:planetmass} shows the disk surface density profiles,  the  positions of the planets and the locations 
of the 2:1 and 3:2 resonances with Jupiter for simulations I1, I3, and I4 just before accretion onto Saturn's core is switched on
As Jupiter grows, we see a clear tendency for Saturn's core to follow the edge of Jupiter's gap where its inward migration (caused by its differential Linblad torque) is balanced by the corotation torque (Masset et al. 2006).  Of particular importance is the location of the gap edge with respect to Jupiter.  In runs I1 and I4 the edge of Jupiter's gap is located just outside the 3:2 resonance whereas for run I3 it lies beyond the 2:1 resonance. 

Fig. \ref{fig:vary_xj} shows the evolution of simulations with different start times for Saturn's accretion ($x_J = 0, 0.5, 1$).
It is interesting to note that Saturn's accretion acts to accelerate Jupiter's accretion (see upper panel).   Jupiter grows to its full mass in $\sim 2900$ and $\sim 1900$ orbits in models I1 and I3, respectively, but takes $\sim 7000$ orbits to acquire its final mass in model I4. This occurs because, as Saturn grows and begins to form a gap, the disk surface density near Jupiter increases, thereby enhancing Jupiter's accretion rate. 

\begin{figure}
\centering
\includegraphics[width=0.95\columnwidth]{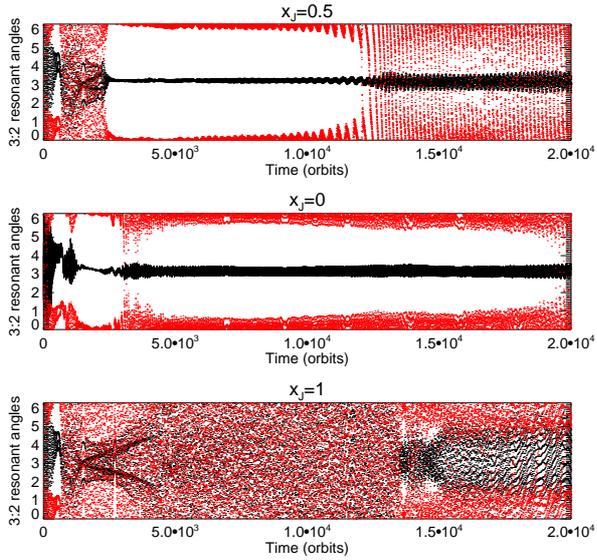}
\caption{For models I1, I3 and I4 the time evolution of the resonant angles 
$\psi=3\lambda_S-2\lambda_J-\omega_J$ (black) and $\phi=3\lambda_S-2\lambda_J-\omega_S$ (red) associated with the 3:2 resonance.}
\label{fig:angles}
\end{figure}

\begin{figure}
\centering
\includegraphics[width=0.8\columnwidth]{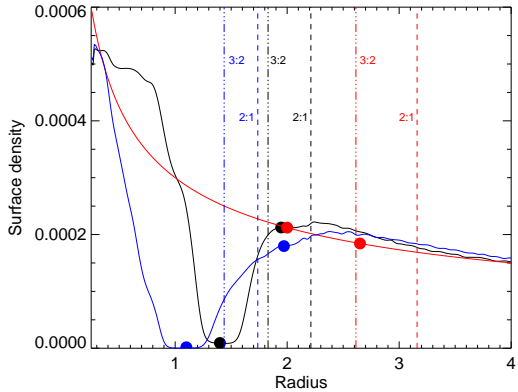}
\caption{The disk surface density 
profile for the three models I1 (black), I3 (red) and I4 (blue) before accretion onto Saturn's 
core is switched on. Also displayed are the locations of the 2:1 resonance (dashed line) and 3:2 
resonance (dot-dashed line) with Jupiter. The dots illustrate the positions of Jupiter and Saturn for the three models. } 
\label{fig:planetmass}
\end{figure}

\begin{figure}
\centering
\includegraphics[width=0.98\columnwidth]{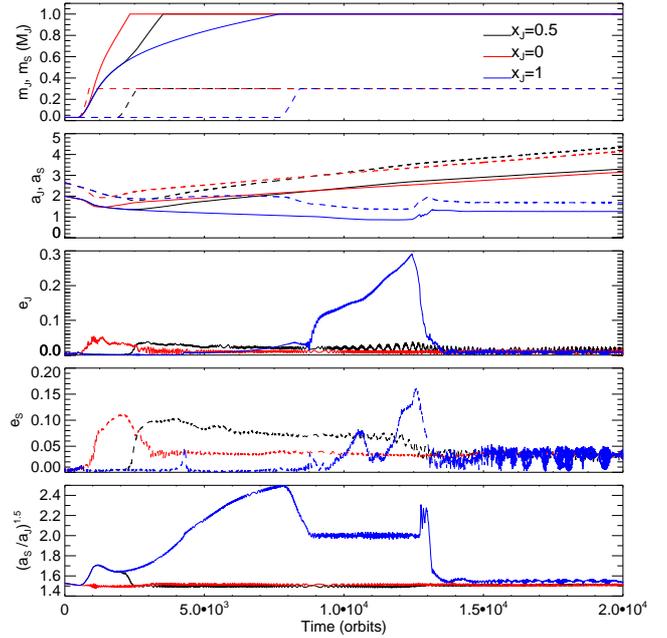}
\caption{{\it Upper (first) panel:}  the evolution of the planet 
masses for models I1 (black), I3 (red) and I4 (blue). {\it Second panel:} evolution of the semimajor axes. 
{\it Third panel:} evolution of Jupiter's 
eccentricity. {\it Fourth panel:} evolution of Saturn's eccentricity. 
{\it Fifth panel:} evolution of the period ratio.}
\label{fig:vary_xj}
\end{figure}

The second panel in Fig. \ref{fig:vary_xj} shows the orbital evolution of Jupiter and Saturn.
The three simulations behave similarly before Jupiter starts to accrete gas from the disk at $t\sim 500$ orbits.  All cases undergo convergent migration of both cores followed by capture in the 3:2 resonance.  
At later times, however, the evolution of the three simulations diverge. For model I3, the fact that gas accretion onto Saturn's core proceeds more rapidly compared with Jupiter leads to an even faster convergent migration compared with model I1, driving 
the planets even deeper into 3:2 resonance after resonant locking. This is illustrated in the second panel of Fig. \ref{fig:angles} which shows, for this simulation, the evolution of the resonant 
angles associated with the 3:2 resonance. In contrast with run I1, periods of circulation of the resonant angles are not observed. Instead, 
gas accretion onto the cores make the libration amplitudes of the resonant angles 
decrease for $t\le 2000$ orbits. 

As before, once Jupiter and Saturn have grown to $m_J\sim 0.5$ $M_J$ and $m_S\sim 0.3$ $M_J$, both planets migrate outward in concert, this time at a slightly faster rate than model I1. 
This is because, as discussed above, in model I3 the faster convergent migration at $t < 2000$ orbits due to Saturn's growth has the consequence of locking the planets more deeply in resonance than in model I1 ( i.e., the libration width of the resonant angles is smaller for I2 than I1; see Fig. \ref{fig:angles}).  At later times however, the amplitude of the resonant angles increases and outward migration slows.  Thus, the final outcome for run I2 is very similar to 
that of model I1 in terms of both the late-time outward migration rate and the planets' eccentricities.

\begin{figure}
\centering
\includegraphics[width=0.95\columnwidth]{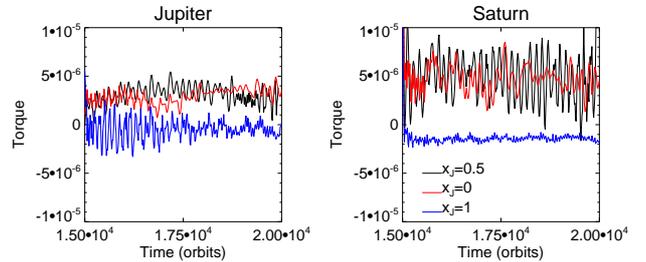}
\caption{Evolution of the torques exerted on Jupiter (left panel) and Saturn (right panel) for 
models I1, I3 and I4 at times where a quasi-stationary state is reached. }
\label{fig:torques}
\end{figure}

\begin{figure}
\centering
\includegraphics[width=0.9\columnwidth]{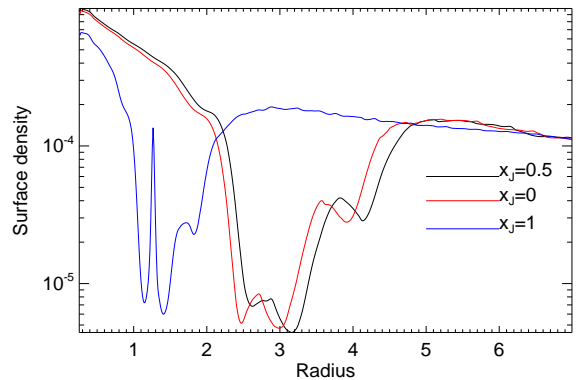}
\caption{The disk surface density profile for models I1 (black), I3 (red) and I4 (blue) at 
$t\sim 1.5\times 10^4$ orbits.}
\label{fig:sigma}
\end{figure}

For simulation I4 -- in which gas accretion onto Saturn's core started only after Jupiter reached its final mass 
-- the evolution differed significantly from runs I1 and 
I3 (Fig. \ref{fig:vary_xj}).
Up to $t\sim 2000$ orbits, which corresponds to $m_j\sim 0.5$ $M_J$, the 
evolution of the system is similar to model I1.  After this point, however, the evolution of run I4 diverges from run I1.  Jupiter type II migrates inward whereas Saturn's 
core migrates slightly outward (second panel in Fig. \ref{fig:vary_xj}).  This is because Saturn's core follows the edge of Jupiter's gap where the positive corotation 
torque balances the negative differential Lindblad torque (Masset et al. 2006).  As Jupiter grows, its gap slowly widens causing the orbital separation between the two planets to increase beyond the $2:1$ resonance (Fig. \ref{fig:planetmass}).  Once Jupiter reaches its final mass,  at $t \sim 7500$ orbits, Saturn's gas accretion starts.  At $t \approx 8000$ orbits, Saturn's growth depletes its coorbital region and the positive corotation torque exerted on Saturn disappears, causing Saturn to once again migrate inward.  As Saturn catches up with Jupiter, 
it is captured in the 2:1 resonance from $t\sim 9000$ to 
$t\sim 1.3\times 10^4$ orbits.  The resonant interaction 
causes significant growth of the planets' eccentricities up to $e_J\sim 0.3$ and 
$e_S\sim 0.15$ (second panel of Fig. \ref{orbits}).  In their high-eccentricity state, the planets briefly repel each other once again outside the 2:1 resonance.  Next, their eccentricities are quickly damped by the disk and Saturn's begins a phase of runaway inward migration (Masset \& Papaloizou 2003).  Its rapid migration allows Saturn to cross over the 2:1 resonance and it is captured in the 3:2 resonance at $t\sim 1.4\times 10^4$ orbits. 

Once trapped in 3:2 resonance, Jupiter and Saturn migrate outward for a short time but, in contrast with models I1 and I3, the outward migration is not maintained. Instead, the planets remain on roughly stationary orbits with 
semimajor axes of $a_J\sim 1.3$ and $a_S\sim 1.7$. 
Fig. \ref{fig:torques} shows the evolution 
of the disk torques for the three models at times when a quasi-stationary state is reached. As 
expected, the disk torque experienced by Jupiter is positive in models I1 and I3 but it oscillates about 
zero in model I4. The total torque exerted on Saturn is clearly positive for 
both models I1 and I3, indicating that the positive corotation torque due to disc material flowing 
from the outer disk across the gap is overcoming the negative differential Lindblad torque in these runs. However, the torque on Saturn is negative in model I4.

\begin{figure}
\centering
\includegraphics[width=0.95\columnwidth]{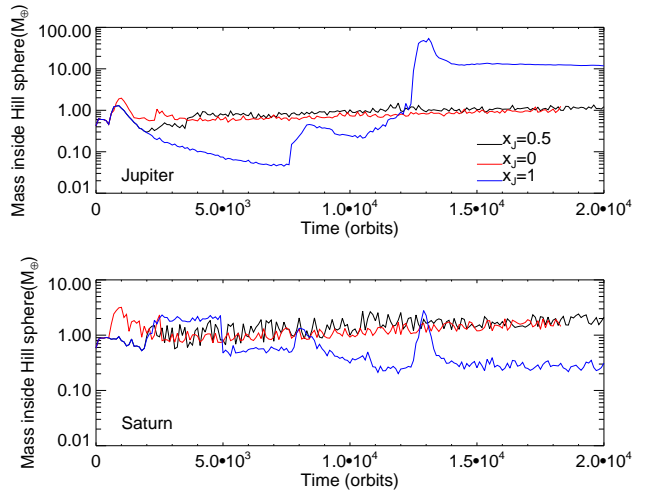}
\caption{ Evolution of the disk mass located inside the Hill sphere of the planets for 
models I1 (black), I3 (red) and I4 (blue).}
\label{fig:hill}
\end{figure}

\begin{figure}
\centering
\includegraphics[width=0.9\columnwidth]{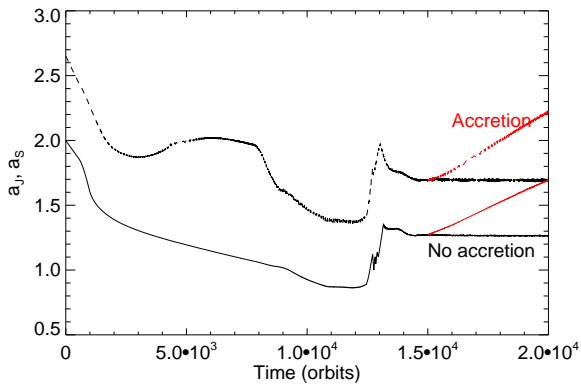}
\caption{ Evolution of the semimajor axes of Jupiter (solid line) and Saturn (dashed line) 
for model I4 (black) and for a restart run in 
which gas is continuously removed from the Hill spheres of the planets (red).}
\label{fig:test_g3}
\end{figure}

To understand the origin of the unexpected zero torque exerted on Jupiter in model I4, Fig. \ref{fig:sigma} shows the disk's surface density profile at $t\sim 1.5\times 10^4$ orbits for the three models. Compared with runs I1 and 
I3, the surface density at the position of Jupiter is much higher in model I4.  Indeed, the amount of disk material enclosed in 
the Hill sphere of the planets is about $\sim1$ $M_\oplus$ in models I1 and I3 and 
$\sim 10$ $M_\oplus$ for model I4.  Fig. \ref{fig:hill} shows that most of the mass in Jupiter's Hill sphere in model I3 was acquired in a short time at 
$t \approx 1.2\times 10^4$ orbits, when Jupiter and Saturn's eccentricities were at their peak and the planets underwent large radial excursions beyond the edges of the gap into the gaseous disk. Although we exclude gas material located 
inside the planet's Hill sphere from the torque calculation, examination of the torque distribution shows that 
the large amount of gas material located in the vicinity of Jupiter's Hill sphere can indeed contribute 
significantly to the total torque exerted on that planet (e.g. Crida et al. 2009). To demonstrate that this disk region is responsible for the stopping of migration, we restarted the run I4 at $t\sim 1.5\times 10^4$ orbits while slowly removing the gas 
within each planets' Hill sphere but while keeping the planets' actual masses fixed. The results of this calculation are shown in Fig. \ref{fig:test_g3} and clearly indicate that outward migration is recovered by removing the residual gas material from the vicinity of Jupiter.  Given that most of the gas bound to Jupiter was acquired at relatively late time, the "correct" outcome of model I4 still includes a complicated orbital evolution (including temporary capture in the 2:1 resonance and eccentricity excitation) but an outward, sustained migration at later times once the planets are trapped in the 3:2 resonance.   The extra gas within Jupiter's Hill sphere, whose inertia was preventing outward migration, should realistically have been accreted or repelled by Jupiter on a relatively short time scale and should not inhibit outward migration.  

Nevertheless, we comment that the high density region located in the vicinity of Jupiter results from the use of an isothermal equation of state, for which pressure gradients correspond to density gradients alone. In non-isothermal disk models, this density peak would be reduced since pressure gradients are partly supported by temperature gradients in that case (Paardekooper \& Mellema 2008). 
 Moreover, in these cases, it is not expected that such a large amount of disk material would accumulate in the planet's vicinity. Indeed, this would
correspond to a large accretion rate onto Jupiter which would result to a significant heating of the
Roche lobe, preventing thereby further accretion (Peplinski et al. 2008).

\subsubsection{Effect of the disk viscosity: model I5}

\begin{figure}
\centering
\includegraphics[width=0.98\columnwidth]{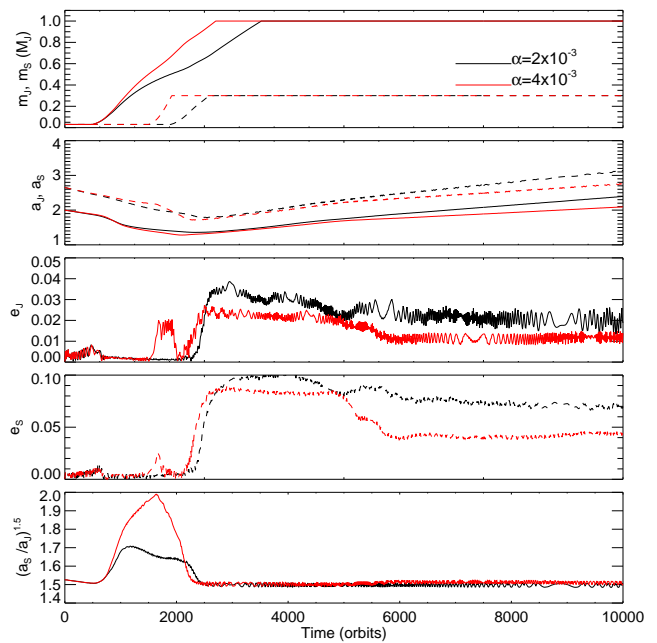}
\caption{{\it Upper (first) panel:} Evolution of the planet masses for models I1 (black) 
and I5 (red). {\it Second panel:} Evolution of the semimajor axes. {\it Third panel:} Evolution of Jupiter's 
eccentricity. {\it Fourth panel:} Evolution of Saturn's eccentricity. 
{\it Fifth panel:} Evolution of the period ratio.}
\label{fig:visc}
\end{figure}

Fig. \ref{fig:visc} illustrates the evolution of a simulation in which $\alpha$ was 
set to $\alpha=4\times 10^{-3}$ (run I5). As a consequence, both the growth and type II migration timescales are shorter.  Indeed, the accretion rates onto the planets are clearly enhanced 
in run I5 compared with run I1 and Jupiter's growth timescale is reduced by $\sim 30\%$. 
Because the viscosity is higher, we expect gap opening to occur later in the growth of Jupiter, i.e., for a higher value of $m_J$.  Using the criterion for gap opening derived by Crida et al. (2006):
\begin{equation}
\frac{3}{4}\frac{H}{R_H}+\frac{50}{q{\cal R}}< 1
\end{equation}
where $q=m_J/M_\odot$, $R_H=a_J(m_J/3M_\odot)^{1/3}$ is the Hill radius of Jupiter and ${\cal R}=a_J^2\Omega_J/\nu$ is the Reynolds number, we indeed predict that gap opening should occur for $m_J>0.23$ $M_J$ in run I1 and for $m_J> 0.35$ $M_J$ in model I5.

Thus, for a higher viscosity, Jupiter's gap grows later and its type II migration is faster.  This means that, when Saturn's gas accretion starts, the Jupiter-Saturn separation is larger for the case of a higher viscosity.  Indeed, for run I5 the two planets are significantly farther apart than for run I1, just interior to the 2:1 resonance.  For even higher viscosities (higher values of 
$\alpha$), Saturn's core would be pushed beyond the 2:1 resonance with Jupiter such that 
subsequent evolution could involve temporary capture in this resonance (as in run I4 discussed above).  

In model I5, the early stages of Saturn's 
growth involve convergent migration of the two planets followed by trapping in the $3:2$ resonance. 
Here, reversal of migration occurs for slightly higher planet masses than seen previously -- $m_J=0.8$ $M_J$ and $m_s=0.3$ $M_J$ 
-- but the final outcome is the same, namely sustained 
outward migration with the planets maintaining their $3:2$ commensurability.  We also find that the outward migration is slower for higher $\alpha$ because Jupiter's gap becomes shallower as the viscosity increases, in agreement with the results of Morbidelli \& Crida (2007).

\subsubsection{Effect of the disk's aspect ratio $h$: models I6, I7}

\begin{figure}
\centering
\includegraphics[width=0.98\columnwidth]{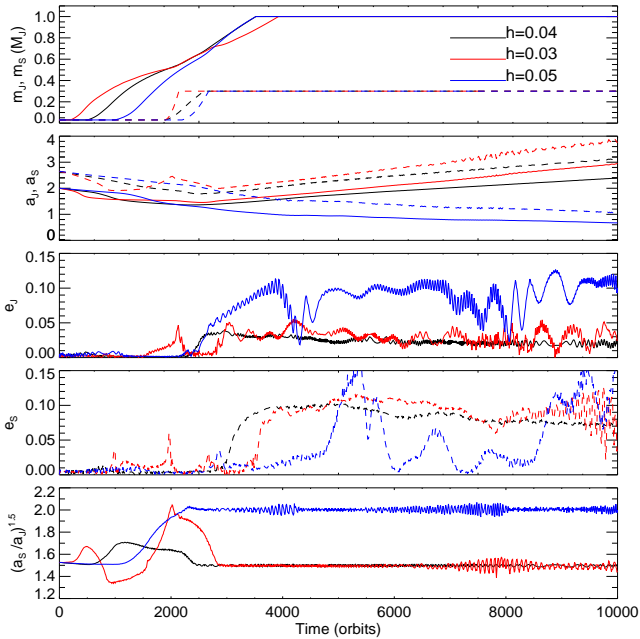}
\caption{{\it Upper (first) panel:}  the evolution of the planet 
masses for models I1 (black), I6 (red) and I7 (blue). {\it Second panel:} evolution of the semimajor axes. 
{\it Third panel:} evolution of Jupiter's 
eccentricity. {\it Fourth panel:} evolution of Saturn's eccentricity. 
{\it Fifth panel:} evolution of the period ratio.}
\label{fig:vary_h}
\end{figure}

We tested the effect of the disk's aspect ratio $h = H/r$ from $h=0.03$ (run I6) to $h=0.05$ (run I7).   Fig. \ref{fig:vary_h} shows the evolution these runs as compared with our fiducial case.  
From Eq. 3 we know that lower-mass planets can open gaps in thinner disks.  Thus, for $h=0.03$, Saturn's core is trapped at the edge of Jupiter's gap early in the simulation, and Saturn's core is pushed outward as Jupiter's mass increases and as its gap widens; this episode of outward migration of Saturn's core is apparent between $1000$ and $2000$ orbits in Fig. \ref{fig:vary_h} ($h=0.03$, second panel). The planets' orbital separation reaches a peak value just outside the 2:1 resonance (see bottom panel of Fig. \ref{fig:vary_h}). For this run ($h=0.03$) accretion onto Saturn's core is switched on at $t\sim 1900$ orbits such that during the early stages of its growth, Saturn still follows Jupiter's gap through the action of the corotation torque. At later times, the interaction with the disk becomes non-linear and Saturn passes through both Jupiter's gap, is captured in the 3:2 resonance 
with Jupiter, and the two planets migrate outward.  Because Jupiter's gap is deeper than in the fiducial case I1, the outward migration is faster for model I6 (see also Morbidelli \& Crida 2007). 

In run I7 the disk was thicker ($h=0.05$) and this resulted in a different mode of evolution.  As before, Saturn's core was captured at the edge of Jupiter's 
gap and pushed outward as the gap widened, this time beyond the 2:1 resonance with Jupiter.   Once Saturn accreted enough gas to cancel the effect of the corotation torque, it became trapped in the 2:1 resonance.  This is because Jupiter's gap is shallower for the thicker disk, causing slower convergent migration of the two planets.  Thus, Saturn is unable to cross the 2:1 resonance (see also Rein et al. 2010).  Of course, disruption of the 2:1 resonance followed by 
capture in the 3:2 resonance on longer timescales can not be ruled out. Indeed, Pierens \& Nelson (2008) showed that, for a scenario close to the setup of model I7, the system is temporarily locked in the 2:1 resonance but the resonance is broken and the planets are evenrually trapped in 3:2 resonance.

\subsubsection{Effect of the disk surface density profile: models I8, I9}

We now test the effect of the disk's radial surface density profile, where the surface density $\Sigma$ varies with orbital radius $R$ as $\Sigma \propto R^{-\sigma}$.  We compare two runs with $\sigma=3/2$ (models I8 and I9) with our standard models that have $\sigma = 1/2$ (I1 and I4).   We note that a $\sigma = 1/2$ profile corresponds to disks with constant accretion rates and 
$\beta=1$, whereas sub-mm measurements of young protoplanetary disks appear to favor $\sigma \approx 0.5-1$ (e.g., Mundy et al. 2000; Andrews \& Williams 2007) and different interpretations of the minimum-mass solar nebula model (Weidenschilling 1977; Hayashi 1981) yields values of $\sigma$ between 1/2 (Davis 2005) and 2 (Desch 2007).  

\begin{figure}
\centering
\includegraphics[width=0.98\columnwidth]{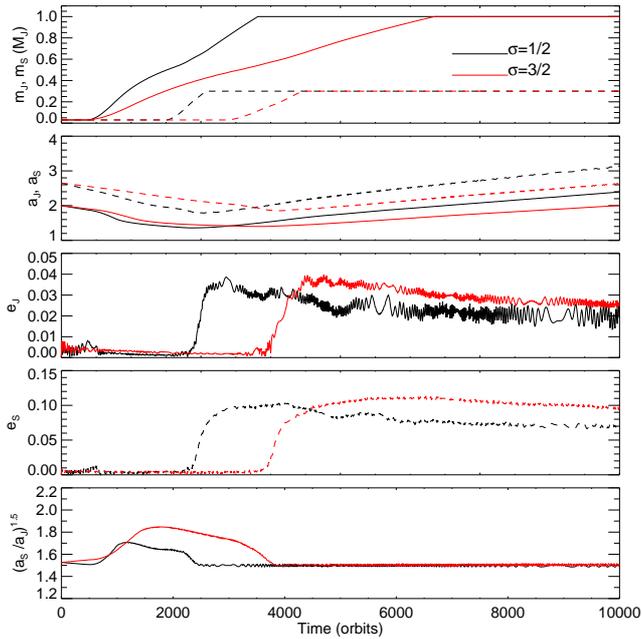}
\caption{{\it Upper (first) panel:}  the evolution of the planet masses
 for models I1 (black) and I8 (red). {\it Second panel:} evolution of the semimajor axes. 
{\it Third panel:} evolution of Jupiter's 
eccentricity. {\it Fourth panel:} evolution of Saturn's eccentricity. 
{\it Fifth panel:} evolution of the period ratio.}
\label{fig:model_g8}
\end{figure}

Eq. \ref {eq:taumig} predicts that for $\sigma=1.5$, Jupiter and Saturn's cores should migrate 
at the same rate (in an isothermal disk) rather than undergoing convergent migration. The consequence is that, in contrast with models in which $\sigma=1/2$, the cores are not locked in 3:2 resonance when Jupiter 
starts to accrete gas from the disk. Fig. \ref{fig:model_g8} shows the evolution of run I8; for this case accretion onto Saturn's core starts when Jupiter has grown to half of its full mass.  In run I9 gas accretion is slower than for model I1 simply because the disk mass included within the feeding zones of the planets is smaller for larger values of $\sigma$.

\begin{figure}
\centering
\includegraphics[width=0.8\columnwidth]{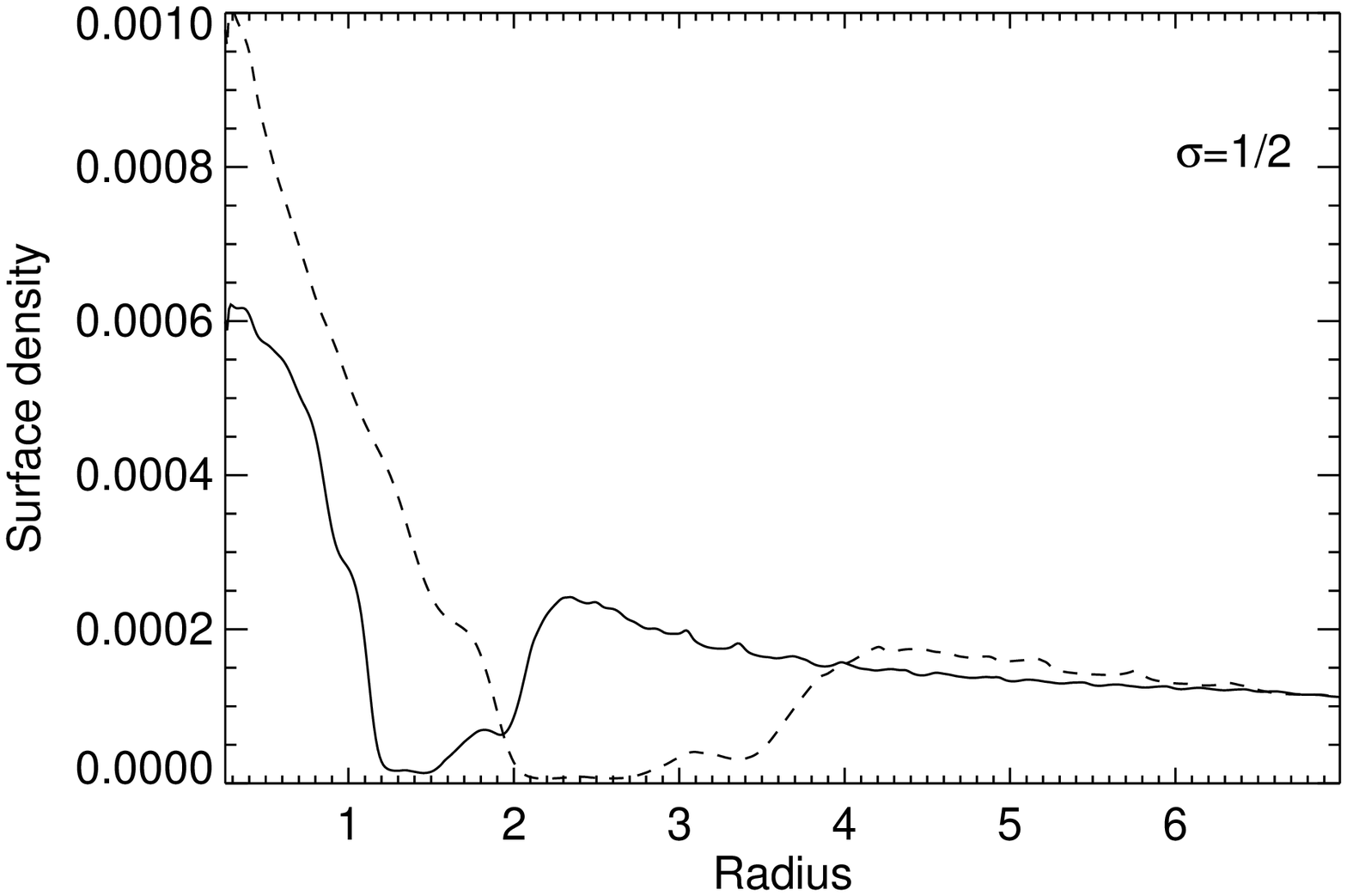}
\includegraphics[width=0.8\columnwidth]{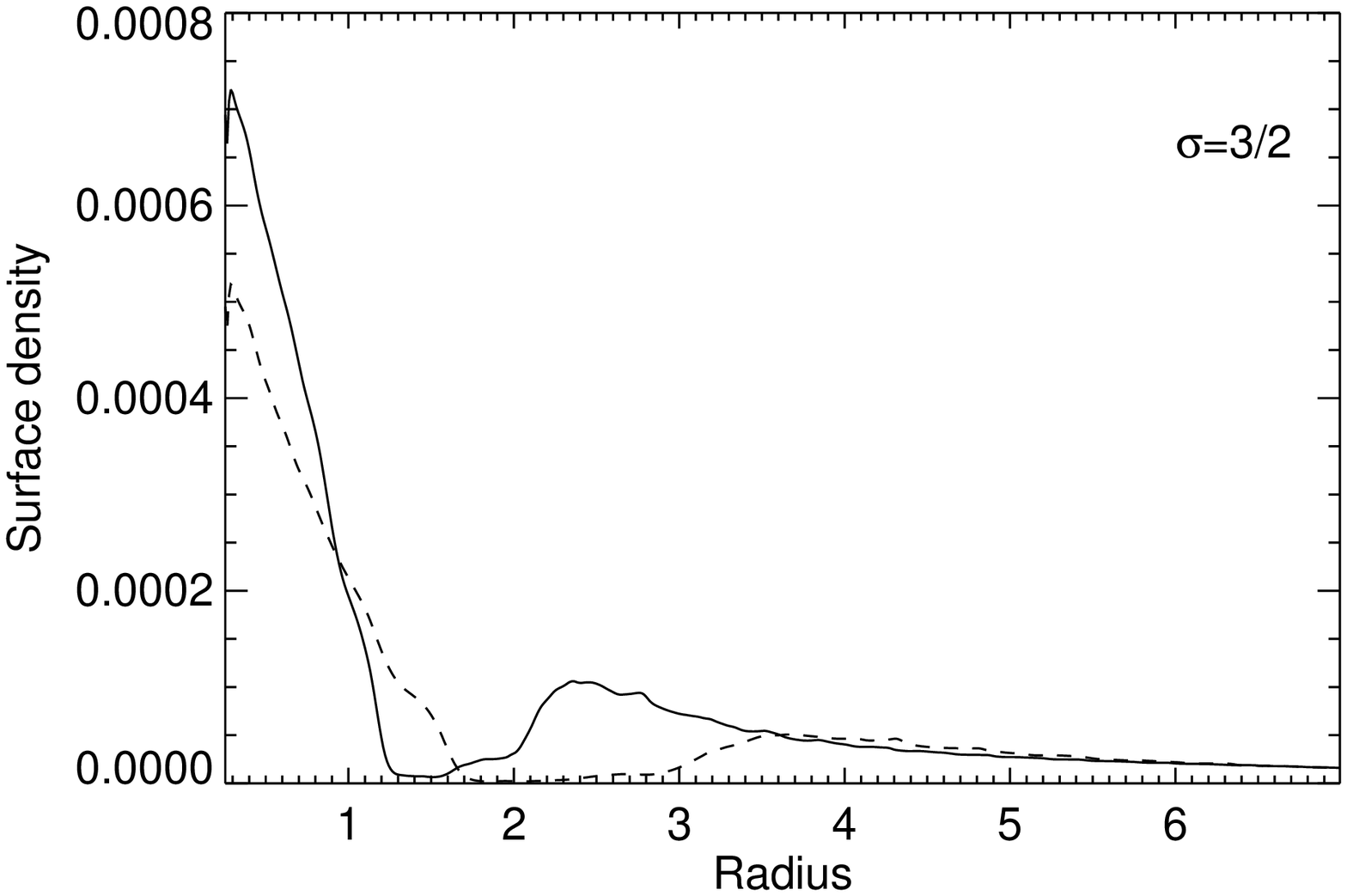}
\caption{{\it Upper panel:}  the disk surface density profile for model I1  prior that outward migration 
of the Jupiter and Saturn system occurs (solid line) and at the end of the simulation (dashed line). {\it Lower panel:} same but for model I8}
\label{fig:sigma_g8}
\end{figure}

The evolution of run I8 is virtually identical to run I1 but slower.  Both the accretion rates and the migration rates are slower for $\sigma=3/2$ (I9). This is simply because of the smaller outer disk mass in a disk with a steep surface density profile; the annular mass scales as $R^{1-\sigma}$ and the mass within a planet's Hill sphere scales as $R^{2-\sigma}$.  Thus, in terms of the outward migration of Jupiter and Saturn, the mass flux across the gap is significantly smaller for the disk with $\sigma = 3/2$. To illustrate this, Fig. \ref{fig:sigma_g8} shows the disk surface density for runs I1 and I8 when outward migration is about to be triggered and at $t=10^4$ orbits. The smaller gas flux across the gap for $\sigma=3/2$ decreases the magnitude of the (positive) corotation torque as well as the density in the inner disk, which has the effect of weakening the (also positive) inner Lindblad torque exerted on Jupiter. This is shown in the time evolution of the disk profiles (Fig. \ref{fig:sigma_g8}): the inner disk surface density increases with time in run I1 due to the gas flowing through the gap whereas 
such an effect is marginal in model I8.

\begin{figure}
\centering
\includegraphics[width=0.98\columnwidth]{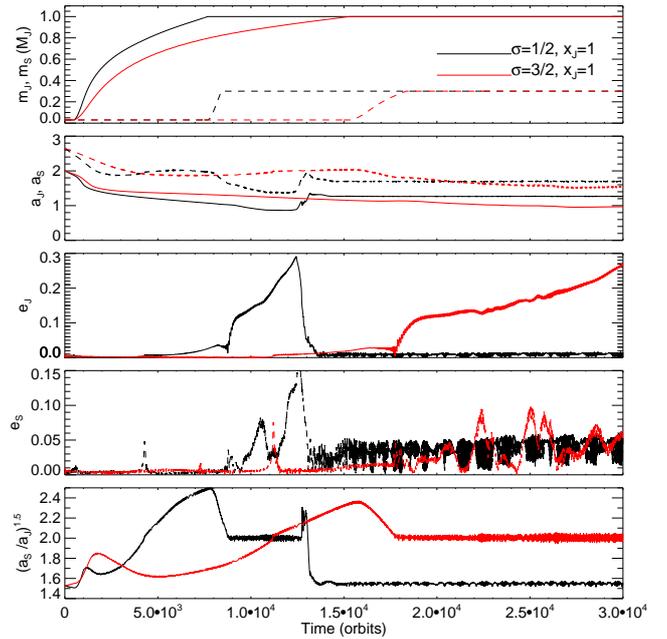}
\caption{{\it Upper (first) panel:}  the evolution of the planet masses for models I1 (black) and I9 (red). {\it Second panel:} evolution of the  semimajor axes. {\it Third panel:} evolution of Jupiter's 
eccentricity. {\it Fourth panel:} evolution of Saturn's eccentricity. 
{\it Fifth panel:} evolution of the period ratio.}
\label{fig:model_g9}
\end{figure}

The disk's surface density profile had only a small impact on the evolution of simulations I1 and I8, in which accretion onto Saturn's core started when Jupiter reached half of its final mass.  This is also the case for runs I4 and I9, in which gas accretion onto Saturn's core is switched on once Jupiter is fully formed (Fig. \ref{fig:model_g9}).  In both runs and as discussed in Section 3.2.1, Jupiter's eccentricity reaches values as high as $e_J\sim 0.3$ during capture in the 2:1 resonance.  And as above, the evolution of the simulation with a steeper disk density profile (run I9) is slower than for the shallower profile (I3).  Thus, although the early evolution of run I9 resembles a stretched-out version of run I3, the final fate of the system is still not reached despite the very long timescale covered by the simulation ($\sim 3\times 10^4$ orbits). However, the fifth panel of  Fig. \ref{fig:model_g9} -- which displays the time evolution of 
the period ratio --  suggests that the 2:1 resonance will be disrupted in the next $\sim 10^4$ orbits. Extrapolating the results from model I1, subsequent evolution should involve capture in 3:2 resonance followed by outward migration of the Jupiter-Saturn system.

\subsubsection{Effects of the gas disk's dispersion} 
\label{sec:disp}

As the Solar Nebula dispersed, the giant planets' migration and accretion stopped.  
We address the effect of the dispersion of gas disk dispersion on the simulation 
presented in Sect. \ref{base} using four additional simulations in which, after a delay, the gas surface density was forced to decay exponentially with an e-folding time $\tau_{disp}$. We used model parameters as in run I1 and we assumed that both the gas disk dispersion and accretion onto Saturn's core start at the same time. 
In these simulations, we varied the value of $\tau_{disp}$ which was set to  $\tau_{disp}=$ $10^3$, $3\times 10^3$ 
and $10^4$ orbits respectively. 

\begin{figure}
\centering
\includegraphics[width=0.98\columnwidth]{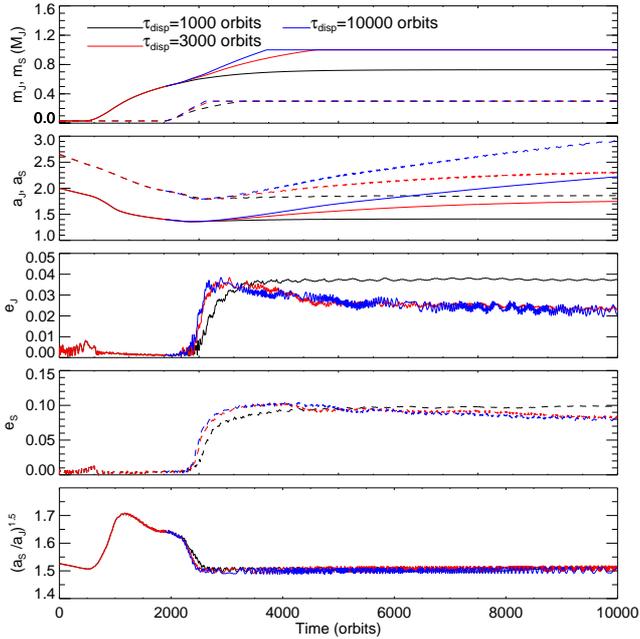}
\caption{{\it Upper (first) panel:}  the evolution of the planet masses for simulations in which 
gas disk dispersion is considered. {\it Second panel:} evolution of the semimajor axes. {\it Third panel:} 
evolution of Jupiter's eccentricity. {\it Fourth panel:} evolution of Saturn's eccentricity. 
{\it Fifth panel:} evolution of the period ratio.}
\label{fig:disp}
\end{figure}

Fig. \ref{fig:disp} shows the evolution of simulations with different values of $\tau_{disp}$. 
The run with $\tau_{disp}=10^3$ orbits is clearly not a viable scenario since the gas lifetime is so short that accretion is cut off early and Jupiter and Saturn never reach their true masses.  For $\tau_{disp}=3\times 10^3$ and $\tau_{disp}=10^4$ orbits, however, the correct masses 
for Jupiter and Saturn are obtained and the system migrates outward, reaching final orbits of $a_J\sim 1.8$ and $a_S\sim 2.4$ for $\tau_{disp}=3\times 10^3$ and $a_J\sim 3.5$ and $a_S\sim 6.5$ for $\tau_{disp}=10^4$ orbits.

\begin{figure}
\centering
\includegraphics[width=0.98\columnwidth]{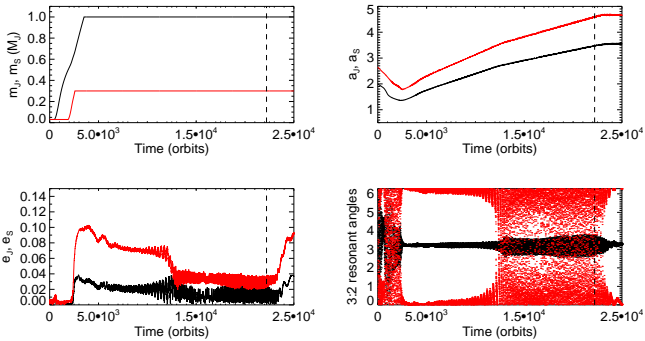}
\caption{{\it Upper left (first) panel:} evolution of the planet masses for 
a simulation with $\tau_{disp}=1000$ orbits and in which onset of disk dispersion 
occurs at $t\sim 2.2\times 10^4$ orbits. {\it Upper right (second) panel:} evolution of the semimajor 
axes. {\it Third panel:} evolution of the eccentricities. {\it Fourth panel:} 
evolution of the period ratio.}
\label{fig:ref}
\end{figure}

The series of simulations presented in this section suggest that a higher value for $\tau_{disp}$ is required 
for such a model to be consistent with the "Grand Tack" scenario.  An alternate possibility is that both 
Jupiter and Saturn formed early in the lifetime of the Solar Nebula, long before the disk was being dispersed. 
To investigate this question, we performed an additional run in which the gas disk disperses when 
Jupiter and Saturn approach their current orbits. Fig. \ref{fig:ref} shows the results of a simulation with 
$\tau_{disp}=1000$ orbits and in which disk dispersion was initiated after $t_{disp}\sim 2.2\times 10^4$ orbits. As expected, the eccentricities grow due to the disk induced eccentricity damping being cancelled and saturate at $e_J\sim 0.05$ and $e_S\sim 0.1$. This effect is also compounded by the fact that the planets become locked 
deeper in the 3:2 resonance while the gas is being dispersed. Here, the planets reach final orbits with 
$a_J\sim 3.6$ and $a_S\sim 4.7$  but it is clear that a similar simulation performed with an adequate value 
for $t_{acc}$ would lead to both Jupiter and Saturn reaching their expected pre-Nice model orbits with $a_J \approx 5.4$ AU (Tsiganis et al. 2005).  

\section{Simulations in radiative disks}

To estimate the influence of a more realistic treatment of the disk thermodynamics, we performed two additional simulations that do not 
use a locally isothermal equation of state but in which the full energy equation is solved (labelled as R1 and R2). 
For run R1, the parameters for the disk and planets were the same as in model I1 whereas for run 
R2, parameters were identical to run I3. 

\begin{figure}
\centering
\includegraphics[width=0.98\columnwidth]{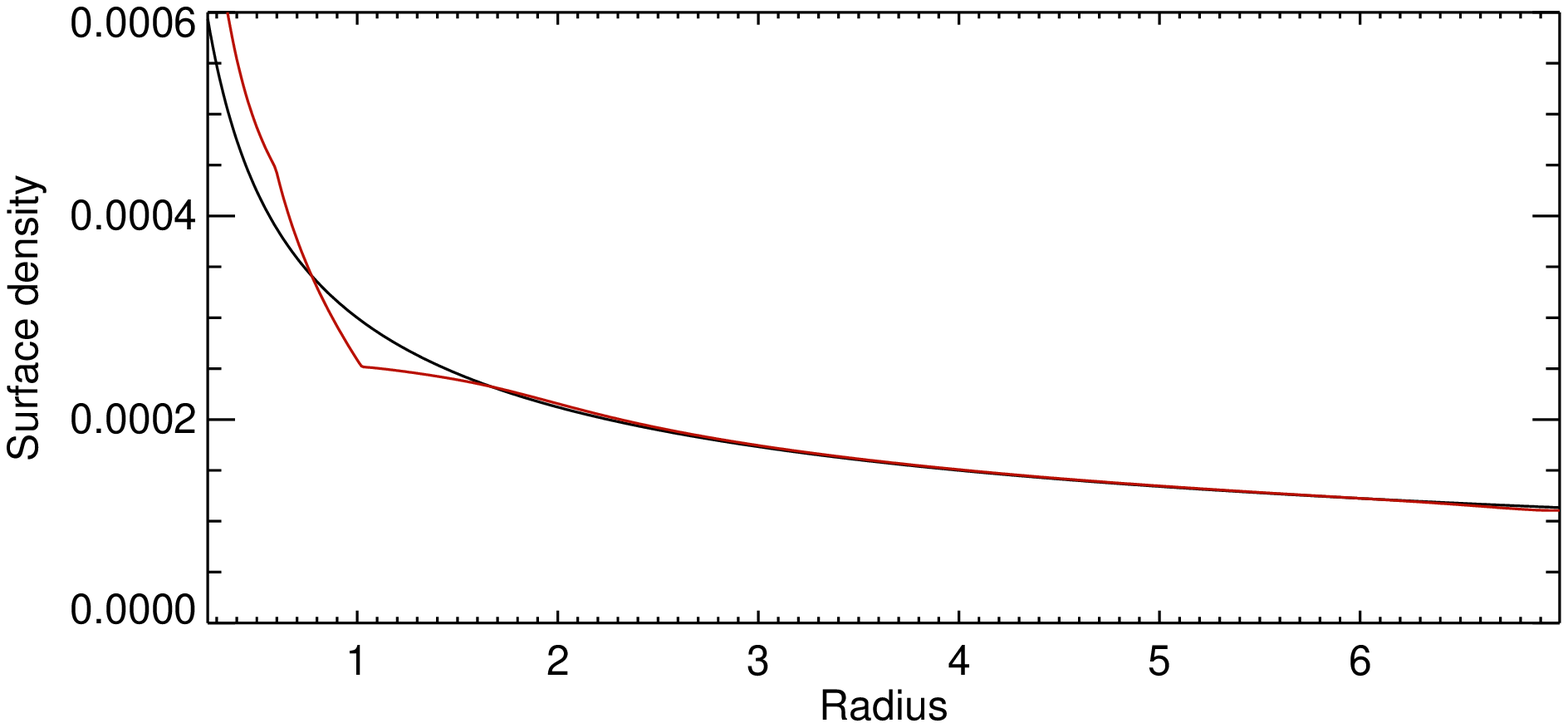}
\includegraphics[width=0.98\columnwidth]{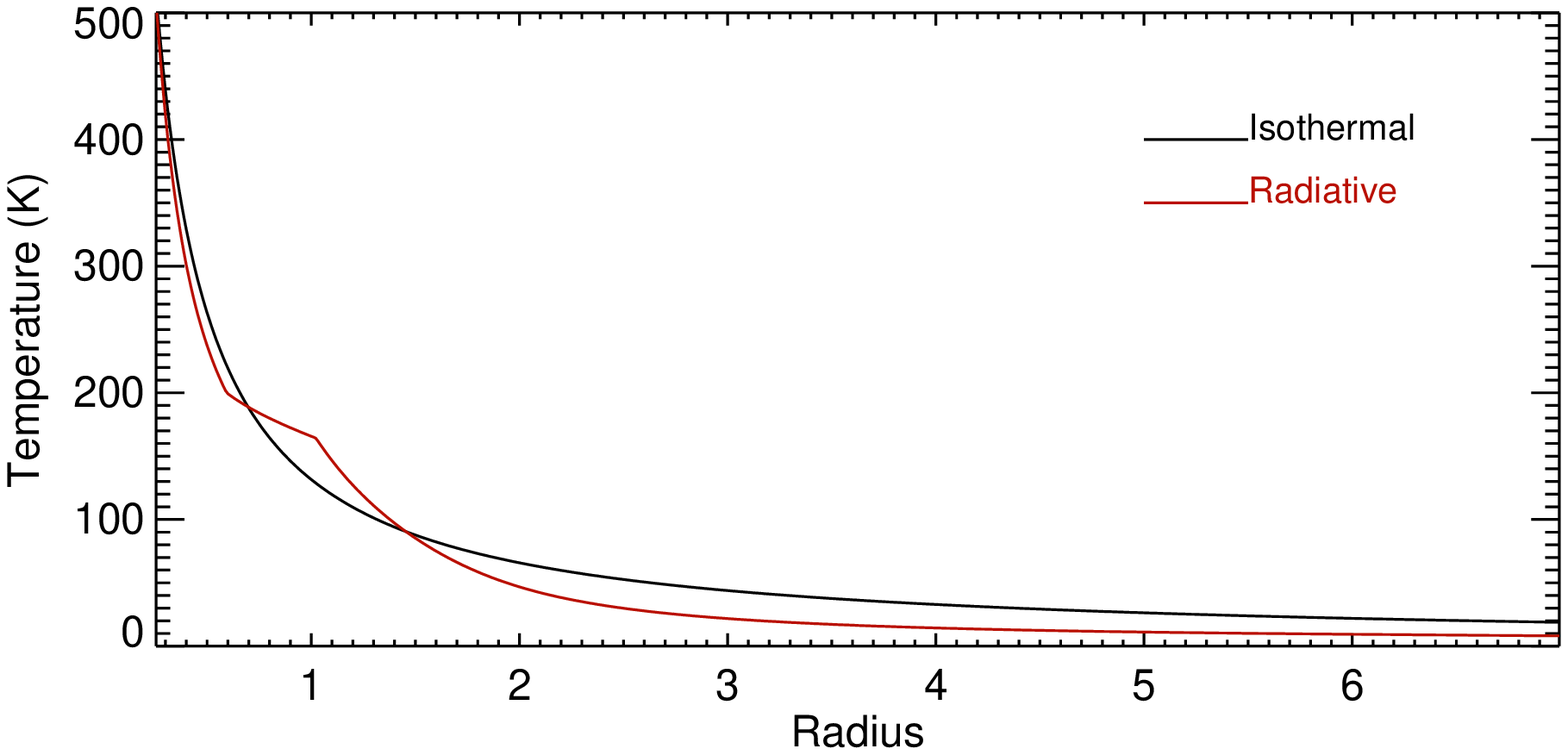}
\caption{ Surface density ({\it upper panel}) and temperature ({\it lower panel}) profiles at equilibrium  for the  
isothermal (black) and radiative (red) models. }
\label{fig:equi_rad}
\end{figure}

For these radiative disk simulations, we ran a preliminary model without any planet in order to obtain a new equilibrium state for the 
disk where viscous heating balances radiative cooling from the disk surfaces. We then restarted the simulation with the cores of Jupiter and Saturn embedded in the disk. Fig. \ref{fig:equi_rad} shows the density and temperature 
profiles when a stationary state for the disk is reached. Although the surface density at the initial positions of Jupiter ($a_J=2$)
and Saturn ($a_S=2.65$) is similar between the isothermal and radiative disks, the temperature at these locations are somewhat
lower in the radiative calculation.  Thus, the disk's aspect ratio is $H/R\sim 3.3\times 10^{-2}$ at the initial position of Jupiter and $H/R\sim 2.8\times 10^{-2}$ at the initial location of Saturn. For the disk model considered here, we note that the disk is initially optically thick for $R\lesssim 6$.

\begin{figure}
\centering
\includegraphics[width=0.98\columnwidth]{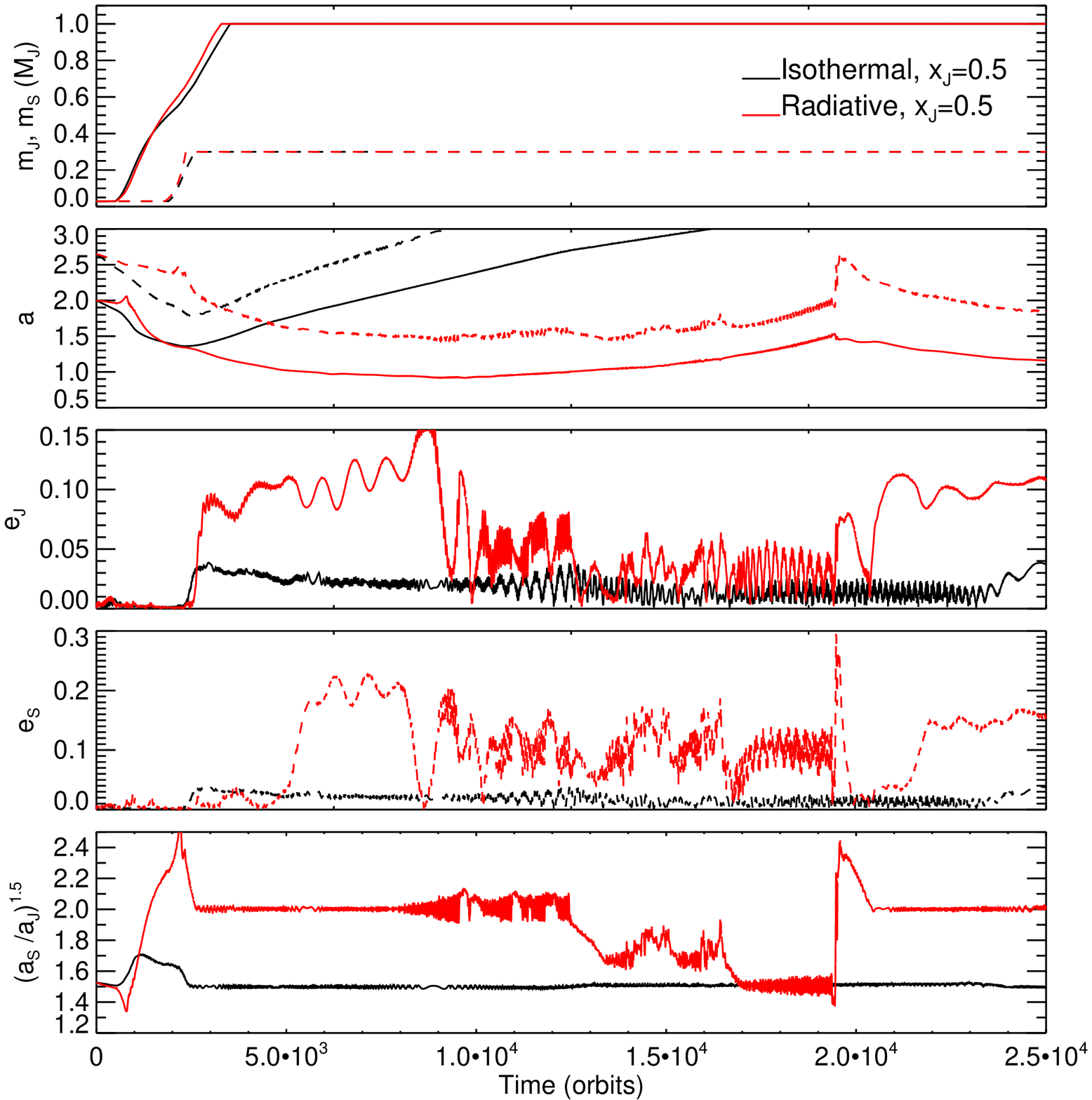}
\caption{{\it Upper (first) panel:}  the evolution of the planet masses for the isothermal model I1 (black) and  the 
radiative model R1 (red). {\it Second panel:} evolution of the  semimajor axes. {\it Third panel:} evolution of Jupiter's 
eccentricity. {\it Fourth panel:} evolution of Saturn's eccentricity. 
{\it Fifth panel:} evolution of the period ratio.}
\label{fig:model_rad}
\end{figure}

Fig. \ref{fig:model_rad} compares the evolution of the system for both the isothermal model I1 and the radiative calculation R1. Although including heating/cooling effects appears 
to have little impact on the mass-growth history of Jupiter and Saturn, the dynamical evolution is very different between the isothermal and radiative disks.  For model R1, the $10$ $M_\oplus$ cores of Jupiter and Saturn initially migrate much more slowly than for run I1. This occurs because the entropy gradient within the horseshoe region of the planets gives rise to a positive 
corotation torque (Baruteau \& Masset 2008; Paardekooper \& Papaloizou 2008) that acts in opposition to the negative differential 
Lindbald torque. This positive corotation torque can be sustained provided that diffusive processes (thermal diffusion, 
heating/cooling effects...) can restore the original temperature profile and that the diffusion timescale across the 
horseshoe region is shorter than the libration timescale $\tau_{lib}$. In the simulations presented here, the diffusion 
timescale corresponds to the vertical 
cooling timescale $\tau_{cool}=c_v\Sigma T/Q$ where $Q$ is the local radiative cooling and $c_v$ the specific heat at constant volume.
For Jupiter, $\tau_{lib}\sim 42$ $T_{orb}$ where $T_{orb}$ is orbital period of the planet and $\tau_{cool}\sim 280$ $T_{orb}$ 
while for Saturn $\tau_{lib}\sim 56$ $T_{orb}$ and $\tau_{cool}\sim 251$ $ T_{orb}$, which means that the corotation 
torque is partially saturated for both embryos. As Jupiter grows, this corotation torque becomes strong enough to push 
the planet outward, which is apparent at $t\sim 1000$ orbits in the second panel of Fig. \ref{fig:model_rad}. When 
Jupiter's mass has reached $\sim 45$ $M_\oplus$ however, the planet opens a gap in the disk which consequently suppresses the 
corotation torque and makes Jupiter migrate inward again on the type II migration timescale. 

Jupiter's type II migration is faster than Saturn's type I migration so the two planets' orbits diverge.  This can be seen for $t\lesssim 2000$ orbits in the fifth panel of Fig. \ref{fig:model_rad}: in this case, the period ratio increased to $(a_S/a_J)^{1.5}\sim 2.4$ before Saturn opened 
a gap in the disk. From this point in time, Jupiter and Saturn migrated convergently until they became captured in the 2:1 resonance. 
The planets then migrated outward for a brief interval but the 2:1 resonance configuration became unstable and was broken at $t\sim 1.2\times 10^4$ orbits. Jupiter and Saturn then became temporarily trapped in 5:3 resonance, during which time slow outward migration continued, but once again the resonance was broken.  Next, Saturn became locked in the 3:2 resonance with Jupiter and the two planets migrated outward together.  However, in contrast with the isothermal runs, this configuration proved unstable, as Jupiter and Saturn underwent a dynamical instability leading to a weak scattering event that launched Saturn beyond the 2:1 resonance at $t\sim 2\times 10^4$ orbits.  At the end of the run, the final fate of the run is still uncertain -- it is possible that the subsequent evolution will again involve a cycle of temporary capture in the 2:1, 5:3 and 3:2 resonances followed by instabilities until the disk dissipates. 

\begin{figure}
\centering
\includegraphics[width=0.98\columnwidth]{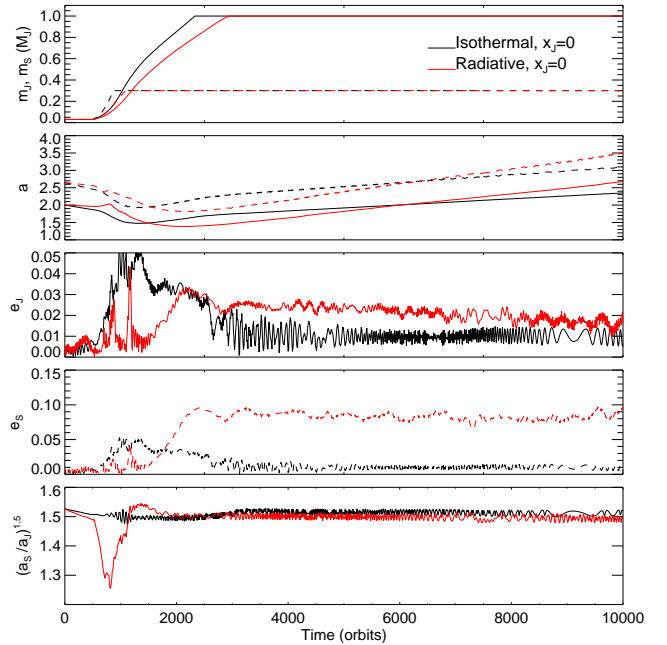}
\caption{{\it Upper (first) panel:}  the evolution of the planet masses for the isothermal model I3 (black) and  the 
radiative model R2 (red). {\it Second panel:} evolution of the  semimajor axes. {\it Third panel:} evolution of Jupiter's 
eccentricity. {\it Fourth panel:} evolution of Saturn's eccentricity. 
{\it Fifth panel:} evolution of the period ratio.}
\label{fig:model_rad2}
\end{figure}

We think that this instability arose because in the outer parts of the radiative disk both the temperature and the disk aspect ratio decrease.  Such a cold disk acts both to accelerate outward migration and to decrease disk-induced eccentricity damping.  Thus, as the two planets migrated outward their eccentricities were significantly higher than for the isothermal runs (see the third and fourth panels of Fig. \ref{fig:model_rad2}).  It is well-known that, for the case of two inward-migrating planets, the inner disk mass plays a key role in damping the inner planet's eccentricity and thus maintaining dynamical stability (Crida et al. 2008).  
Because the disk aspect ratio is smaller, the gap edge here lies further from the planets compared with isothermal runs, resulting in a weaker eccentricty 
damping from the inner disk.  Simulation R1 appears to present a similar scenario but with two outward-migrating planets, with Saturn's relatively large and chaotically-varying eccentricity acting as the trigger for instability.  
It should be noted that the outer parts of the radiative disk are so cold because viscosity is the only heating process in the simulation.  However, it is well known that stellar irradiation is the main heat source at $R>3$ AU (D'alessio et al. 1998).  Thus, a more realistic disk should probably have a warmer outer disk. It is unclear if this would encourage outward migration by reducing the likelihood of instability or discourage outward migration by overly puffing up the disk.  This is an area for future study.  

Fig. \ref{fig:model_rad2} show the evolution of radiative run R2 in which Jupiter and Saturn start to accrete 
gas at the same time. As it grows, Jupiter's outward migration due to the entropy-related corotation torque makes the planets converge, until their orbital period ratio reaches a minimum of $\sim 1.25$ at $t\sim 800$ orbits (i.e., Saturn is interior to the 3:2 resonance).  Once it has accreted enough gas to cancel the effect of the corotation torque, 
Jupiter migrates inward again, resulting in a divergent migration which continues until the planets become locked 
in the 3:2 resonance.  Outward migration of Jupiter and Saturn is then triggered and appears to be maintained until the simulation was stopped after $10^4$ orbits.  Compared with the isothermal disk, Saturn's eccentricity is significantly higher although its behavior is steady and not obviously chaotic as in run R1.  We do not know if this outward migration will continue indefinitely or whether the system might be subject to an instability similar to run R1.   

Thus, simulations R1 and R2 demonstrate that periods of outward migration of Jupiter and Saturn in radiative disks are viable.  However, only one of two simulations produced a clear two-phase migration.  Given the limitations in our simulations (especially with regards to the thermal state of the outer disk), we do not know whether a two-phase migration of Jupiter and Saturn is a likely outcome in radiative disks.  Indeed, these radiative simulations were performed assuming that the radius $R=1$ in the 
computational domain corresponds to $5$ AU. Contrary to isothermal runs, it is worth to note that results from radiative simulations can not be 
scaled to apply to different parameters. Unless the Grand Tack occured in the last stages of the disk's lifetime, our radiative
calculations  
therefore probably underestimate the disk temperature at the location where Jupiter's 
migration reversed.  We are currently working to test the outward migration mechanism of Masset \& Snellgrove (2001) in more realistic radiative disks under a range of physical conditions.  


\section{Evolution of the Solar Nebula}

Our results thus far show that a two-phase migration of Jupiter and Saturn is extremely robust in isothermal disks but is as-yet uncertain in radiative disks.  Protoplanetary disks can be considered to be isothermal if they are optically thin (i.e., if the optical depth $\tau < 1$) and radiative if they are optically thick ($\tau > 1$).  But when in the Solar Nebula's history was it isothermal or radiative? 

To address this question we constructed a simple, 1-D model of the viscously-evolving Solar Nebula.  The disk extended from 0.1 to 40 AU and initially contained $40$ $M_J$ following an $R^{-1/2}$ surface density profile.  We adopted an $\alpha$ prescription for the disk's viscosity  (Shakura \& Sunyaev 1973) and used the same value as in most of the hydro simulations, $\alpha = 2\times 10^{-3}$. We solved the viscous diffusion for the surface density:
\begin{equation}
\frac{\partial \Sigma}{\partial t}=\frac{3}{R}\frac{\partial}{\partial R}\left[\sqrt{R}\frac{\partial \sqrt{R} \nu\Sigma }{\partial R}\right]
\end{equation}
and calculated the temperature using a simple radiative balance between the disk's viscous heating and radiative cooling (as in Lyra et al 2010): 
\begin{equation}
2 \sigma T^4 = \tau_{eff} \left(\frac{9}{4} \nu \Sigma \Omega^2 \right),
\end{equation}
\noindent where $\sigma$ is the Stephan-Boltzmann constant and $\Omega$ is the orbital frequency.  The effective optical depth to the disk midplane is represented by $\tau_{eff}$, which is defined as
\begin{equation}
\tau_{eff} = \frac{3\tau}{8} + \frac{\sqrt{3}}{4} + \frac{1}{4 \tau}.
\end{equation}
The optical depth is $\tau = \kappa \Sigma /2$.  We assume that the opacity $\kappa$ is dominated by small grains and use the values from Bell \& Lin (1994). 

\begin{figure}
\centering
\includegraphics[width=0.98\columnwidth]{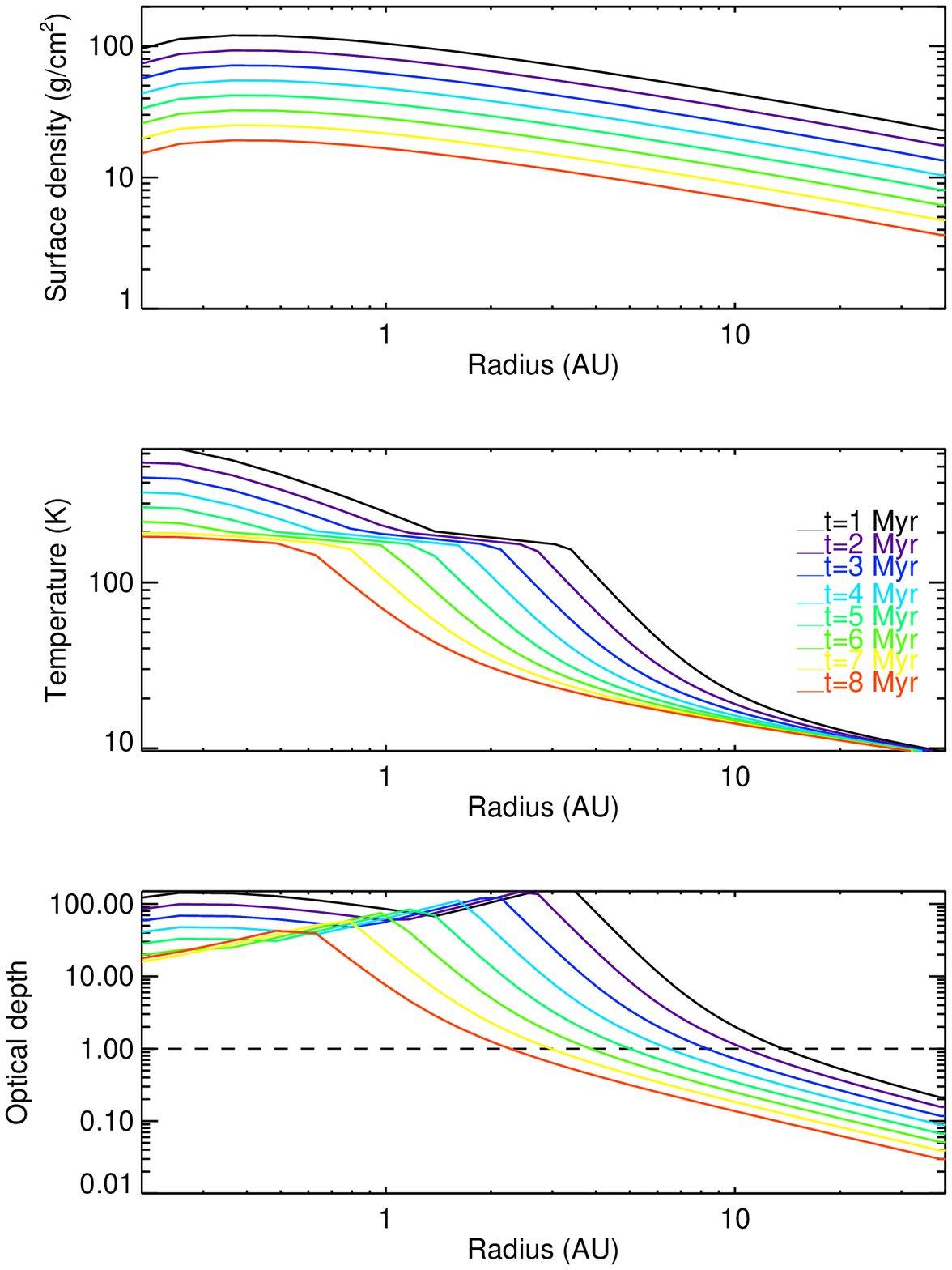}
\caption{{\it Upper (first) panel:}  Surface density profiles for the one-dimensional disk model from $t=1$ Myr (upper 
curve) to $t=8$ Myr.    {\it Second panel:} Evolution of the temperature. {\it Third panel:} Evolution of the optical depth.}
\label{fig:1dmodel}
\end{figure}

Fig.~\ref{fig:1dmodel} shows 8 Myr in the evolution of a representative Solar Nebula.  As the disk viscously spreads its surface density decreases uniformly and the disk cools.  The cooling is not uniform due to the large variations in opacity between different temperature regimes.  Similarly, the disk is initially optically thick in its inner 20 AU and optically thin farther out.  In time, the boundary between optically thick and thin moves inward but interior to 1 AU the disk remains optically thick throughout.  Because we have not included photo-evaporation, the disk's density continues to decrease but never to zero.  In reality, at some point we expect the disk to be completely removed by either photo-evaporation (Hollenbach et al 1994; Adams et al. 2004) or perhaps an MRI-related instability (Chiang \& Murray-Clay 2007).  The optical depth in a given location depends on the local disk properties, although the evolution of the disk's surface density profile certainly depends on whether the disk dissipates from the outside-in or the inside-out.  

How does the Grand Tack fit in the context of this simple model?  The region of interest, from roughly 1-10 AU, is clearly in the radiative regime early in the disk's lifetime.  In the last few Myr this region transitions to an isothermal state.  The relevant boundary where an isothermal disk should affect Jupiter and Saturn's evolution is the outer edge of Saturn's gap when Saturn is at its closest to the Sun.  This corresponds to about 2.5 AU for Jupiter at 1.5 AU and Saturn in 2:3 MMR at 1.97 AU.  At 2.5 AU, the disk transitions from radiative to isothermal after roughly 6 Myr of evolution, although the time at which this state is reached is parameter-dependent.  

When the disk transitions to an isothermal state its density is significantly decreased compared to its initial state or to the configuration of the hydrodynamical simulations presented in Sects. 3 and 4.  Could Jupiter and Saturn migrate from 1.5-2 AU out to their current locations in such a low-mass disk?  There are two criteria that should be required to allow for long-range outward migration via the Masset \& Snellgrove (2001) mechanism.  First, the inner lindblad torque acting on Jupiter must be larger than the outer lindblad torque acting on Saturn.  Second, the angular momentum content of the gas though which Jupiter and Saturn will migrate must be sufficient to transport them a long distance.  At the time of the radiative-to-isothermal transition the torque balance criterion is met if we assume that the disk profile at that time should have been sculpted by Jupiter and Saturn as in the hydrodynamical simulations.  In addition, the angular momentum content in the  gas from 2.6 to roughly 8-10 AU is $\sim 1.5$ times larger than that needed to move Jupiter and Saturn to 5.4 and 7.1 AU.

Despite meeting the theoretical criteria for outward migration,  we ran two additional hydrodynamical simulations to see if outward migration of Jupiter and Saturn could truly occur in such low-mass disks. One simulation was run with an isothermal equation of state whereas the other included radiative effects.  Jupiter and Saturn started the simulations fully-formed and were placed just exterior to the 3:2 resonance.  The disk mass interior to Jupiter's orbit was only $\sim 0.4 M_J$.In addition, in these two simulations the spatial units were AU rather than multiples of 5 AU (and the corresponding time units years rather than $5^{3/2} \approx 10$ years) such that these simulations truly test the Grand Tack at its correct scale.  Given the large computational expense these simulations were only run for 1000 years and serve mainly as a proof of concept. 

\begin{figure}
\centering
\includegraphics[width=0.98\columnwidth]{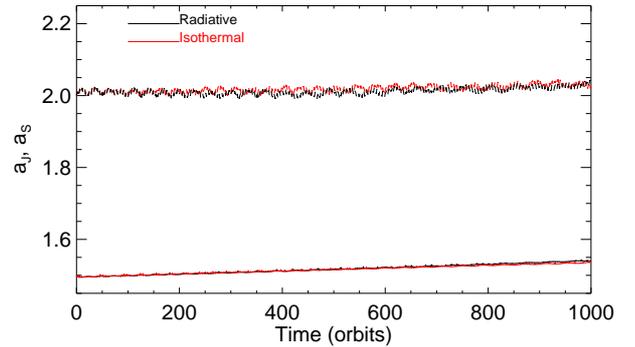}
\caption{Time evolution of the semimajor axes of Jupiter (lower line) and Saturn (upper line) for hydrodynamical simulations adapted to the 
Solar Nebula.}
\label{fig:run_sn}
\end{figure}

Fig. \ref{fig:run_sn} shows the time evolution of the semimajor axes of Jupiter and Saturn in these two simulations.  As predicted, Saturn became trapped in resonance and the two planets tacked and migrated outward in both cases.  This confirms the results found in previous sections and shows that the Grand Tack mechanism applies on the relevant spatial scale.  

It therefore appears that outward migration of Jupiter and Saturn from 1.5-2 AU to beyond 5 AU is a natural outcome in an isothermal Solar Nebula.  Of course, in certain situations outward migration may occur in a radiative disk (e.g., Fig~\ref{fig:model_rad}).  But, for the limiting case in which outward migration can only occur in an isothermal disk, we still expect the Grand Tack to happen because in the last stages of the disk's lifetime it is necessarily optically thin and, as seen in Sect. 3, this leads inevitably to outward migration.  

\section{Summary and Discussion}

Our results indicate that Jupiter and Saturn probably underwent a two-phase, inward-then-outward migration.  In our simulations, Jupiter and Saturn start as $10$ $\mearth$ cores and type I migrate; inward for isothermal disks, inward or outward for radiative disks.  In most cases the two cores become locked in 3:2 mean motion resonance (MMR).  At this point  or after a delay of 500-1000 orbits, we allowed Jupiter to start accreting gas from the disk.  When Jupiter reaches the gap-opening mass, it undergoes a phase of rapid inward migration as it clears out its gap (sometimes called type III migration; Masset \& Papaloizou 2003) then settles into standard, type II migration.  Inward migration continues until Saturn accretes enough gas to reach the gap-opening mass itself.  At this point, Saturn's inward migration accelerates and is again trapped in the 3:2 MMR with Jupiter.  Outward migration of both giant planets is then triggered via the mechanism of Masset \& Snellgrove (2001). Outward migration stops when either a) the disk dissipates (as in Sect. \ref{sec:disp}), b) Saturn reaches the outer edge of the disk, or, c) if the disk is flared, the giant planets drop below the local gap-opening mass (e.g., Crida et al. 2009).  

 An additional stopping -- or at least slowing -- mechanism exists if the planets are unable to maintain a well-aligned resonant lock during migration.  For example, in simulation I1 Jupiter and Saturn's rate of outward migration slowed significantly when one resonant angle transitioned from libration to circulation (see Figs. 1 and 3).  Here, this appears to be due to a positive corotation torque exerted on Saturn by gas that polluted Jupiter and Saturn's common gap as the distance between the two planets increased during outward migration.  On longer timescales it is unclear if this mechanism would continue to slow down and eventually stop the outward migration.

In isothermal disks, the two phase migration of Jupiter and Saturn holds for almost the full range of parameters that we tested (Sect. 3).  The only situation for which this result does not hold is if the gaseous Solar Nebula is relatively thick ($h = H/r \gtrsim 0.05$).  Both the disk's surface density profile and the value for the disk aspect ratio had an effect on the migration rate: disks with either shallower profiles or lower values of $h$ result in faster migration. Changing the disk's viscosity had little effect on the outcome, although we only tested a very small range.  In one simulation (I4; Sect. \ref{sec:varyxj}), Saturn's core was pushed past the 2:1 MMR with Jupiter leading to a significant eccentricity increase for both planets before the resonance was crossed, Saturn was trapped in the 2:3 MMR and both planets migrated outward.  This dynamic phase of resonance crossing and eccentricity excitation is likely to be quite sensitive to the detailed properties of the disk (e.g., the scale height and viscosity) that determine the gap profile. 

We performed two simulations in radiative disks with mixed outcomes (Sect. 4).  In the first case, Jupiter and Saturn started accreting together and so stayed relatively close to each other.  The planets became locked in the 3:2 MMR and migrated outward even faster than in isothermal simulations due to the small aspect ratio of the outer disk ($h \approx 0.03$).  In the second case, Saturn started to accrete when Jupiter reached half its final mass (i.e., $x_J = 0.5$), by which time the two planets were beyond the 2:1 MMR.  The planets succeeded in breaking the 2:1 and 5:3 MMRs, became trapped in the 3:2 MMR and migrated outward only to undergo a dynamical instability putting the planets once again beyond the 2:1 MMR.  A more detailed study of outward migration in radiative disks is underway.  

Using a simple 1-D model of an evolving Solar Nebula we showed that the disk should be optically thick at early times, then transition to optically thin from the outside-in during the late phases of its evolution.  At the orbital distance in question (1-10 AU), the disk transitions from radiative to isothermal behavior in the last 1-2 Myr of its evolution.  Thus, even if we make the ``pessimistic'' assumption that an isothermal disk is required for outward migration of Jupiter and Saturn, the disk fulfills the criteria for long-range outward migration in its late phases.  Outward migration of Jupiter and Saturn at this time is very likely provided the disk remains thin ($h \lesssim 0.05$).

Our simulations therefore show that an inward-then-outward migration of Jupiter and Saturn is extremely likely, and that the last phase of outward migration probably coincided with the late phases of the dissipation of the Solar Nebula.  This is of particular interest because the two phase migration of Jupiter and Saturn helps resolve a long-standing problem in terrestrial planet formation.  For over 20 years, simulations of terrestrial accretion have been unable to reproduce Mars' relatively small mass ($0.11$ $\mearth$; Wetherill 1978, 1991; Chambers 2001; Raymond et al. 2009).  This problem arises because, in a Solar Nebula that varies smoothly in orbital radius, there is a comparable or larger amount of mass in the vicinity of Mars than the Earth.  For Mars to be so much smaller than Earth, most of the mass between roughly 1-3 AU must be removed (e.g., Raymond et al. 2006, 2009; O'Brien et al. 2006).  Several mechanisms have been proposed to remove this mass, including strong secular resonances (Thommes et al. 2008, Raymond et al. 2009) and a narrow dip in the surface density caused by a radial dependence of the disk's viscosity (i.e., a dead zone; Jin et al. 2008).  However, the problem is most easily and much better solved if the terrestrial planets did not form from a wide disk of planetary embryos but instead from a narrow annulus extending only from 0.7-1 AU (Wetherill 1978; Chambers 2001; Hansen 2009).  In that case, Mars' small mass is simply an edge effect: Mars is small was built from  one or perhaps a few embryos that were scattered beyond the edge of the embryo disk (this is also the case for Mercury, which was scattered inward beyond the inner edge of the embryo disk).  In contrast, Earth and Venus formed within the annulus and are consequently much more massive.  Simulations of terrestrial planet formation can quantitatively reproduce the orbits and masses of all four terrestrial planets as well as their radial distribution (Hansen 2009).  

The flaw in simulations of terrestrial planet formation in truncated disks is that they had no justification for the truncation; the ad-hoc initial conditions were simply chosen because they provided a good fit to the actual terrestrial planets (Hansen 2009).  The two phase migration of Jupiter and Saturn provides such a justification via the Grand Tack model of Walsh et al. (2011).  If Jupiter's turnaround point was at $\sim 1.5$ AU then it would have naturally truncated the inner disk of embryos and planetesimals at about 1 AU -- in most of our simulations Jupiter indeed tacked at roughly this distance.  As expected, the terrestrial planets that form from this disk quantitatively reproduce the actual terrestrial planets (Walsh et al. 2011).  The Grand Tack model also provides the best explanation to date for the observed dichotomy between the inner and outer asteroid belt (Gradie \& Tedesco 1982).  Thus, the present-day Solar System appears to bear the imprint of a two phase migration of Jupiter and Saturn.   Our hydrodynamical simulations provide support for the Grand Tack scenario.  

As with any numerical study, our simulations do not fully represent reality.  The aspect of our simulations that is probably the least realistic is the gas accretion onto the giant planets' cores.  
In our simulations, gas accretion onto Jupiter and Saturn is extremely fast.  Once accretion starts, Jupiter and Saturn reach their final masses in only a few thousand years, whereas the Kelvin-Helmholtz time in protoplanetary disks is more like $\sim 10^5$ years.  In addition, accretion onto growing giant planet cores requires transferring gas through circum-planetary accretion disks whose physical properties are poorly constrained (e.g., Ward \& Canup 2010).  Once the planets reached their actual masses we artificially turned off gas accretion. If, during the outward migration Saturn accreted enough gas to carve a gap as deep as Jupiter's then Saturn's outer lindblad torque would balance Jupiter's inner lindblad torque, outward migration would stop and the planets would turn back around and migrate inward.  The impact of a more realistic accretion history on Jupiter and Saturn's migration remains an open question, in particular with regards to the interplay between gas accretion and the dispersal of both the cimcumstellar and cicumplanetary disks.

\begin{acknowledgements}{}
We thank Alessandro Morbidelli, Kevin Walsh and Franck Selsis for helpful
discussions. Some of the simulations were performed using HPC resources from GENCI-cines (c2011026735). 
We are grateful to the CNRS's PNP and EPOV programs and to the Conseil
Regional d'Aquitaine for their funding and support.  
\end{acknowledgements}

\end{document}